\documentclass[a4paper,10pt,twoside]{cpc-hepnp}

\usepackage{CJK,upgreek,fancyhdr}
\usepackage{subfigure}
\usepackage{multicol}
\usepackage{units}
\usepackage{booktabs}
\usepackage{amsmath}
\usepackage{amsfonts,amssymb,bm,mathrsfs,bbm,amscd}
\usepackage{lastpage}
\usepackage{multirow}
\usepackage{stfloats}
\usepackage{graphicx}
\DeclareGraphicsExtensions{.pdf,.jpeg,.png}
\usepackage{epstopdf}
\usepackage[english]{babel}
\usepackage{overpic}
\usepackage[title]{appendix}
\usepackage{lineno}
\usepackage{color}
\usepackage{diagbox}
\usepackage{epsfig}
\usepackage{bigstrut}
\usepackage{overpic}
\usepackage{dcolumn}
\usepackage{xspace}
\usepackage{verbatim}
\usepackage{footmisc}
\usepackage{xcolor}
\usepackage{soul}

\usepackage{caption}
\usepackage{float}

\usepackage[colorlinks,linkcolor=blue,anchorcolor=blue,citecolor=blue]{hyperref}

\hypersetup{hypertex=true,
	colorlinks=true,
	linkcolor=blue,
	anchorcolor=blue,
	citecolor=blue}

\newcommand{\ifb}{\mathrm{fb}^{-1}}

\newcommand{\ee}{e^+e^-}
\newcommand{\pip}{\pi^+}
\newcommand{\pim}{\pi^-}
\newcommand{\mev}{\,\mathrm{MeV}}

\newcommand{\mevcc}{\,\mathrm{MeV}/c^2}
\newcommand{\gev}{\,\mathrm{GeV}}
\newcommand{\gevc}{\,\mathrm{GeV}/c}

\newcommand{\Modea}{\bar{p}{}K_{S}^0}
\newcommand{\Modeb}{\bar{p}{}K^{+}\pi^-}
\newcommand{\Modec}{\bar{p}{}K_{S}^0\pi^0}
\newcommand{\Moded}{\bar{p}{}K_{S}^0\pi^-\pi^+}
\newcommand{\Modee}{\bar{p}{}K^{+}\pi^-\pi^0}
\newcommand{\Modef}{\bar{p}{}\pi^-\pi^+}
\newcommand{\Modeaa}{\bar{\Lambda}{}\pi^-}
\newcommand{\Modebb}{\bar{\Lambda}{}\pi^-\pi^0}
\newcommand{\Modedd}{\bar{\Lambda}{}\pi^-\pi^+\pi^-}
\newcommand{\Modeaaa}{\bar{\Sigma}{}^{0}\pi^-}
\newcommand{\Modeccc}{\bar{\Sigma}{}^{-}\pi^0}
\newcommand{\Modeddd}{\bar{\Sigma}{}^{-}\pi^-\pi^+}

\newcommand{\lambdacp}{\Lambda_{c}^{+}}

\newcommand{\ie}{\emph{i}.\emph{e}.}

\newcolumntype{d}[1]{D{.}{.}{#1}}

\begin{document}
\begin{CJK*}{UTF8}{gkai}

\fancyhead[c]{\small Chinese Physics C~~~Vol. xx, No. x (2022) xxxxxx}

\fancyfoot[C]{\small 010201-\thepage}
\footnotetext[0]{Received xxxx June xxxx}

\title{Observations of the  Cabibbo-Suppressed decays $\Lambda_{c}^{+}\to n\pi^{+}\pi^{0}$, $n\pi^{+}\pi^{-}\pi^{+}$ and the  Cabibbo-Favored decay  $\Lambda_{c}^{+}\to nK^{-}\pi^{+}\pi^{+}$\thanks{
This work is supported in part by National Key R$\&$D Program of China under Contracts Nos. 2020YFA0406400, 2020YFA0406300; National Natural Science Foundation of China (NSFC) under Contracts Nos. 11635010, 11735014, 11805086, 11835012, 11935015, 11935016, 11935018, 11975011, 11961141012, 12022510, 12025502, 12035009, 12035013, 12192260, 12192261, 12192262, 12192263, 12192264, 12192265; the Chinese Academy of Sciences (CAS) Large-Scale Scientific Facility Program; Joint Large-Scale Scientific Facility Funds of the NSFC and CAS under Contract No. U1832207; the CAS Center for Excellence in Particle Physics (CCEPP); 100 Talents Program of CAS; Fundamental Research Funds for the Central Universities, Lanzhou University, University of Chinese Academy of Sciences; The Institute of Nuclear and Particle Physics (INPAC) and Shanghai Key Laboratory for Particle Physics and Cosmology; ERC under Contract No. 758462; German Research Foundation DFG under Contracts Nos. 443159800, Collaborative Research Center CRC 1044, GRK 2149; Istituto Nazionale di Fisica Nucleare, Italy; Ministry of Development of Turkey under Contract No. DPT2006K-120470; National Science and Technology fund; National Science Research and Innovation Fund (NSRF) via the Program Management Unit for Human Resources $\&$ Institutional Development, Research and Innovation under Contract No. B16F640076; STFC (United Kingdom); Suranaree University of Technology (SUT), Thailand Science Research and Innovation (TSRI), and National Science Research and Innovation Fund (NSRF) under Contract No. 160355; The Royal Society, UK under Contracts Nos. DH140054, DH160214; The Swedish Research Council; U. S. Department of Energy under Contract No. DE-FG02-05ER41374.}}
\maketitle
\begin{center} 
\begin{small}
	\begin{center}
		M.~Ablikim(麦迪娜)$^{1}$, M.~N.~Achasov$^{11,b}$, P.~Adlarson$^{70}$, M.~Albrecht$^{4}$, R.~Aliberti$^{31}$, A.~Amoroso$^{69A,69C}$, M.~R.~An(安美儒)$^{35}$, Q.~An(安琪)$^{53,66}$, Y.~Bai(白羽)$^{52}$, O.~Bakina$^{32}$
% ...
		\\
		\vspace{0.2cm}
		(BESIII Collaboration)\\
		\vspace{0.2cm} {\it
			$^{1}$ Institute of High Energy Physics, Beijing 100049, People's Republic of China\\
% ...
$^{4}$ Bochum Ruhr-University, D-44780 Bochum, Germany\\
% ...
$^{11}$ G.I. Budker Institute of Nuclear Physics SB RAS (BINP), Novosibirsk 630090, Russia\\
% ...
$^{31}$ Johannes Gutenberg University of Mainz, Johann-Joachim-Becher-Weg 45, D-55099 Mainz, Germany\\
			$^{32}$ Joint Institute for Nuclear Research, 141980 Dubna, Moscow region, Russia\\
% ...
$^{35}$ Liaoning Normal University, Dalian 116029, People's Republic of China\\
% ...
$^{52}$ Southeast University, Nanjing 211100, People's Republic of China\\
			$^{53}$ State Key Laboratory of Particle Detection and Electronics, Beijing 100049, Hefei 230026, People's Republic of China\\
% ...
$^{66}$ University of Science and Technology of China, Hefei 230026, People's Republic of China\\
% ...
$^{69}$ (A)University of Turin, I-10125, Turin, Italy; (B)University of Eastern Piedmont, I-15121, Alessandria, Italy; (C)INFN, I-10125, Turin, Italy\\
			$^{70}$ Uppsala University, Box 516, SE-75120 Uppsala, Sweden\\
% ...
$^{b}$ Also at the Novosibirsk State University, Novosibirsk, 630090, Russia\\
% ...
	}\end{center}
	
	\vspace{0.4cm}
\end{small}

\end{center}

\begin{abstract}
Using electron-positron annihilation data samples corresponding to an integrated luminosity of 4.5\,\unit{fb}$^{-1}$, collected by the BESIII detector in the energy region between $4599.53~\mev$ and $4698.82~\mev$, we report the first observations of the Cabibbo-suppressed decays $\Lambda_c^+\to n\pi^+\pi^0$, $\Lambda_c^+\to n\pi^+\pi^-\pi^+$, and the Cabibbo-favored decay $\Lambda_c^+\to nK^-\pi^+\pi^+$ with statistical significances of $7.9\sigma$, $7.8\sigma$, and $>10\sigma$, respectively.
The branching fractions of these decays are measured to be $\mathcal{B}(\Lambda_{c}^{+}\rightarrow n\pi^{+}\pi^{0})=(0.64\pm0.09\pm0.02)\%$, $\mathcal{B}(\Lambda_{c}^{+}\rightarrow n\pi^{+}\pi^{-}\pi^{+})=(0.45\pm0.07\pm0.03)\%$, and  $\mathcal{B}(\Lambda_{c}^{+}\rightarrow nK^{-}\pi^{+}\pi^{+})=(1.90\pm0.08\pm0.09)\%$, where the first uncertainties are statistical and the second are systematic.
We find that the branching fraction of the decay  $\Lambda_{c}^{+}\rightarrow n\pi^{+}\pi^{0}$ is about one order of magnitude higher than that of $\Lambda_{c}^{+}\rightarrow n\pi^{+}$.
\end{abstract}



\begin{multicols}{2}

\section{Introduction}
Charmed baryons provide an excellent laboratory for studying the properties of QCD in the case where a heavy quark couples to two light quarks.
Currently, we do not have reliable phenomenological models to describe these complex baryon decays.
More than two decades ago, a general formulation of a topological-diagram scheme for the nonleptonic weak decays of baryons was proposed to calculate the amplitudes of different topological diagrams~\cite{Chau:1995gk}.
The factorizable external $W$-emission amplitude $T$ and internal $W$-emission amplitude $C$, as well as the non-factorizable inner $W$-emission amplitude $C^{\prime}$ and $W$-exchange amplitude $E$ are introduced in the representation of these  diagrams.
In this scheme, one amplitude can be factorized into two parts, the decay constant of the emitted meson and the heavy-to-light transition form factor, which can be directly calculated.
By way of example, the topological diagrams for $\Lambda_{c}^{+}\to n\pi^{+}\pi^{0}$ can be seen in Fig.~\ref{fig:feynman}.
We know that here the non-factorizable contribution is essential for understanding the weak decays of charmed baryons, in contrast to the negligible effect they contribute in heavy-meson decays~\cite{Cheng2021}.
Hence, studies of charmed baryon decays involving non-factorizable components are critical for understanding the underlying dynamics of charmed-baryon decays in general.
\begin{center}
	\centering
	 \begin{overpic}[width=0.22\textwidth]{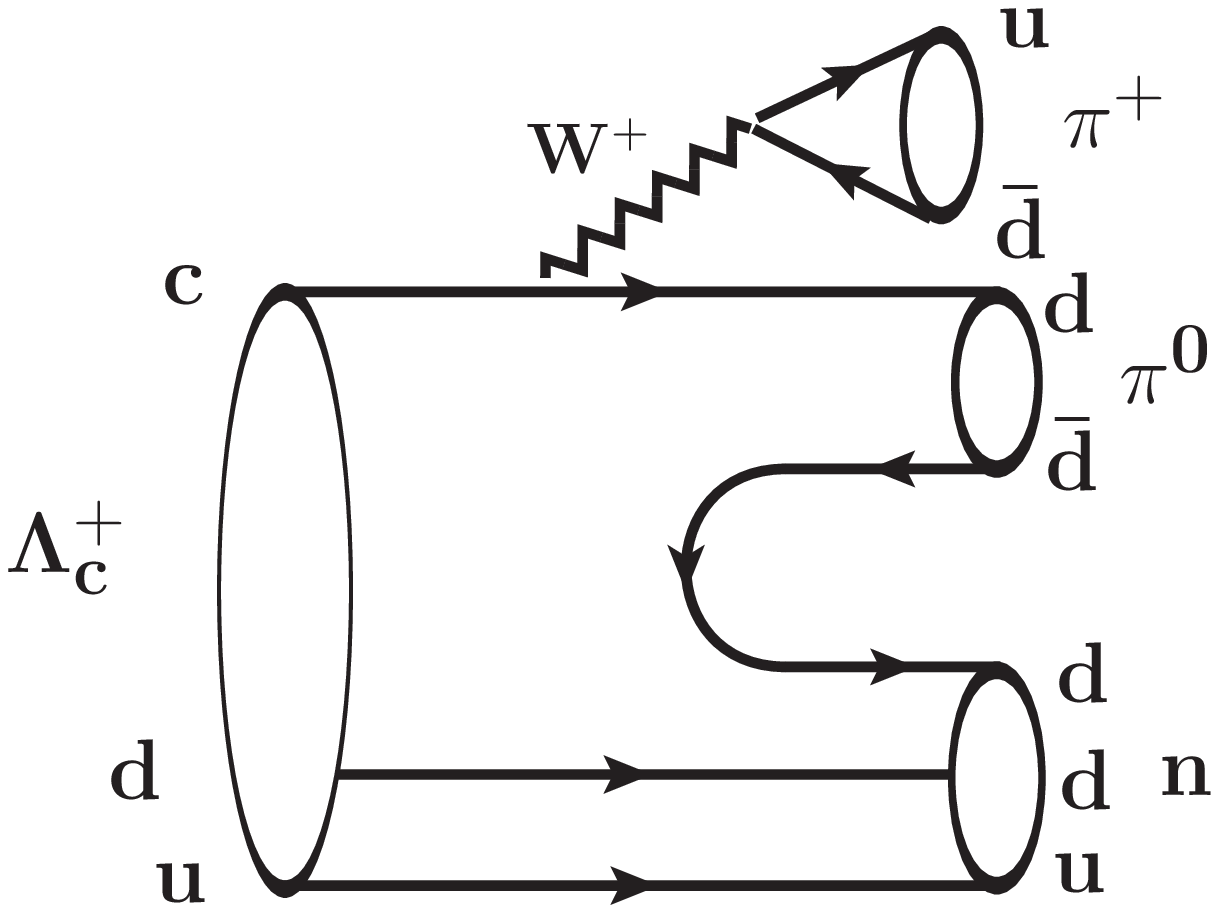}
		\put(25, 60){\scriptsize{(a)}}
		\put(50, -10){\scriptsize{$T$}}
	\end{overpic}
	\begin{overpic}[width=0.22\textwidth]{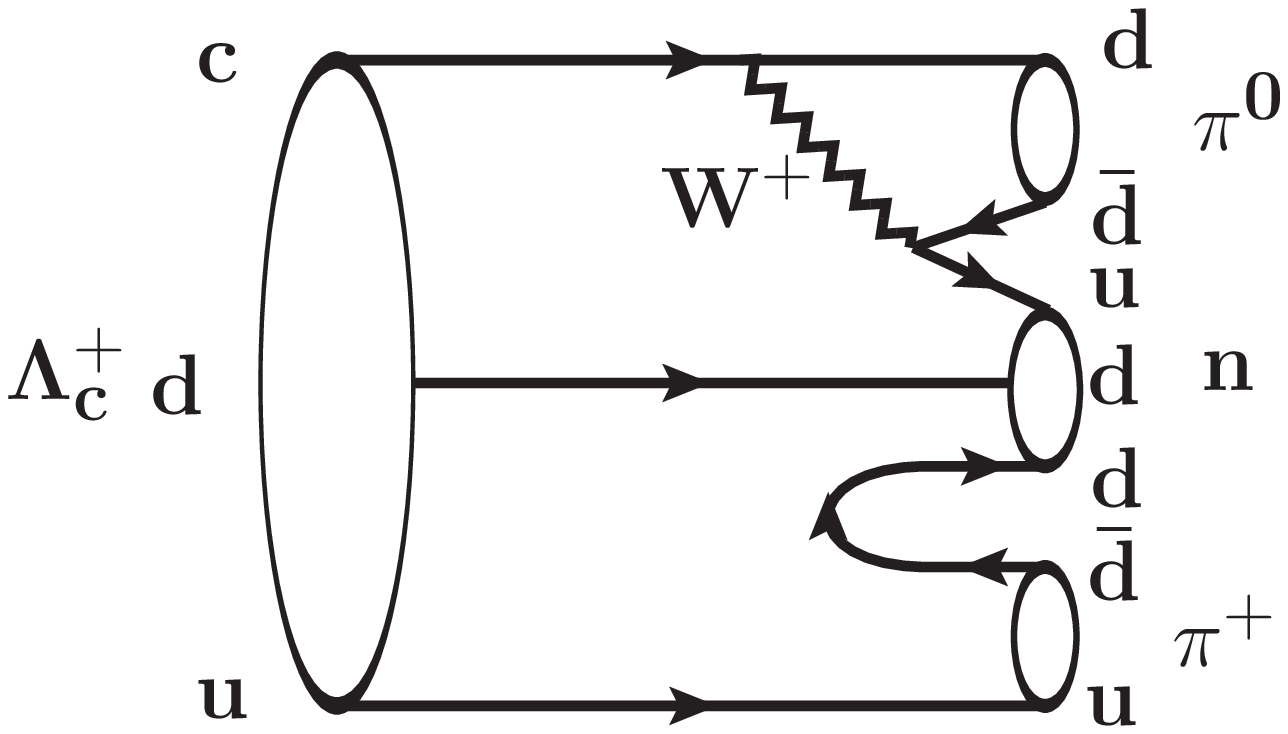}
		\put(25, 60){\scriptsize{(b)}}
		\put(50, -10){\scriptsize{$C$}}
	\end{overpic}
	\\ \vspace{0.8cm}
	\begin{overpic}[width=0.22\textwidth]{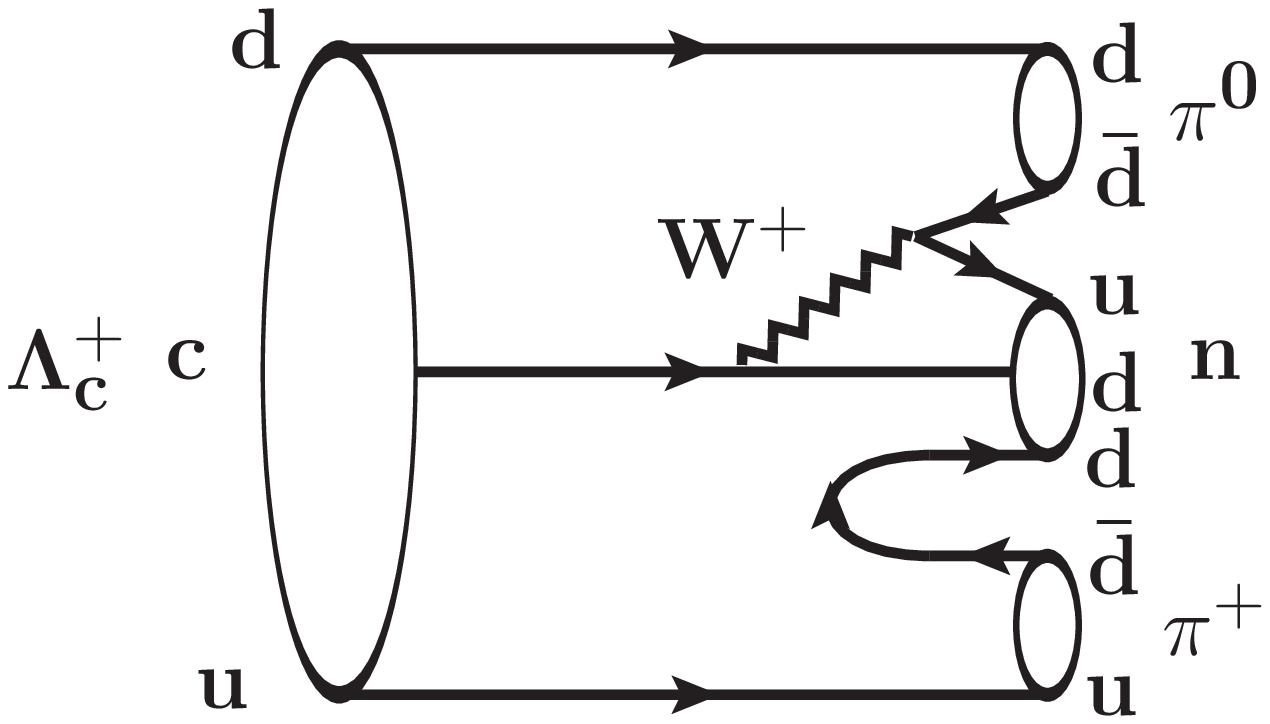}
		\put(25, 60){\scriptsize{(c)}}
		\put(50, -10){\scriptsize{$C^{\prime}$}}
	\end{overpic}
	\begin{overpic}[width=0.22\textwidth]{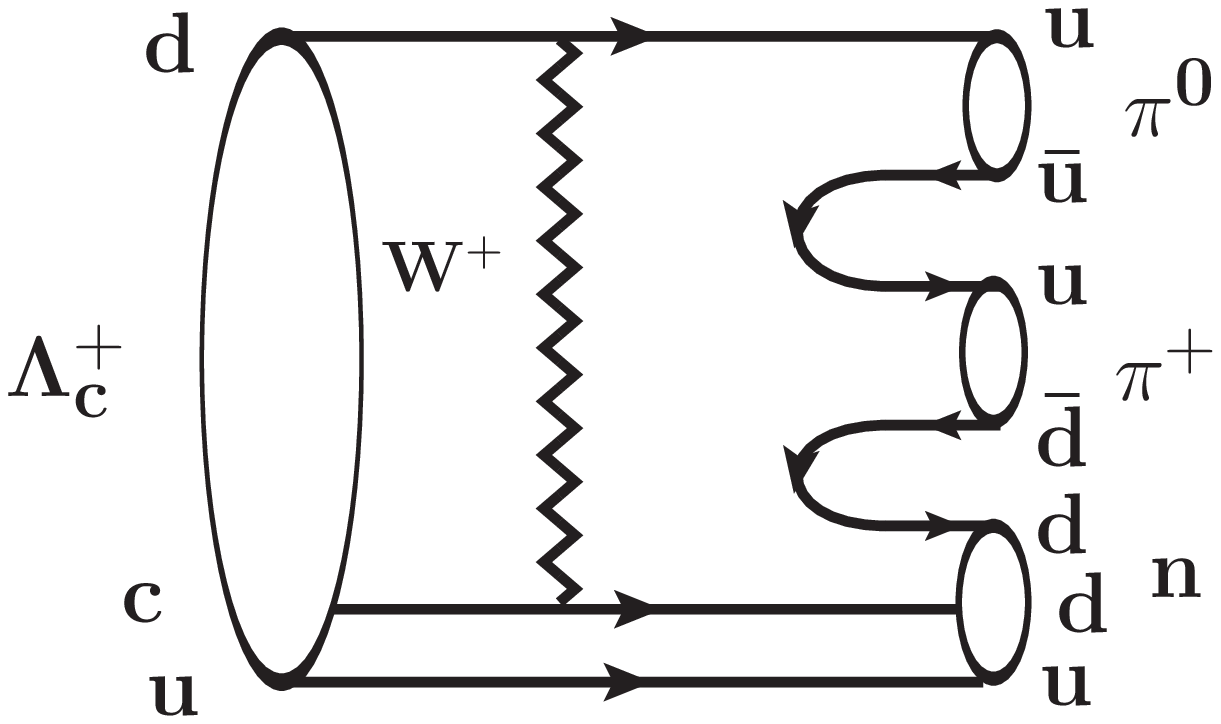}
		\put(25, 60){\scriptsize{(d)}}
		\put(50, -10){\scriptsize{$E$}}
	\end{overpic}  
	\vspace{0.32cm}
	\figcaption{Topological diagrams of   the decay $\Lambda_c^+ \to n\pi^+\pi^0$ via (a) external $W$-emission $T$, (b) internal $W$-emission $C$, (c) inner $W$-emission $C^{\prime}$, and (d) $W$-exchange diagram $E$.}
	\label{fig:feynman}
\end{center}

In addition, an alternative and model-inpendent approach based on quark flavor SU(3) symmetry has been proposed to describe charmed-baryon decays. Even though it is an approximate method, it is a powerful and reliable tool to extract useful information about these transitions~\cite{Wang2016, Geng2017, Geng2018, Geng2018_2, Geng2019, Geng2019_2, He2018, Xiao2020}.
Improved measurements of charmed-baryon  decays will be important in further testing the validity  of this method.

Many experimental studies of various $\Lambda^+_c$ decays have been reported, with a focus on decays to  two-body hadronic final states~\cite{Li:2021iwf, BESIII:hadron, BESIII:ppi0peta, BESIII:nkspi, BESIII:xik, BESIII:sigmaeta,  BESIII:Lmdpieta,  BESIII:npi, LHCb:pomega, Belle:ppi0peta,  Belle:pomega}.
However, more studies of multi-body hadronic decays  are required to gain deeper insight into the nature of non-perturbative QCD, as these involve rich intermediate processes and the branching fractions (BF) are in general higher.
There is a significant discrepancy between the predicted BFs of $\Lambda_c^+\to N \bar{K} \pi\pi$ decays reported in Ref.~\cite{Statmodel2020}  and the measured values for  $\Lambda_c^+\to pK^-\pi^+\pi^0$ and $\Lambda_c^+\to p\bar{K}^0\pi^+\pi^-$~\cite{pdg2021}.
Clarifying this tension requires knowledge of the BFs of hadronic $\Lambda_c^+$ decays to final states involving a neutron. Fewer studies have been performed for these decays because of the difficulty of direct neutron detection.

The BESIII collaboration has measured the BFs of the Cabibbo-favored (CF) decay $\Lambda_{c}^{+}\rightarrow nK_{S}^{0}\pi^{+}$~\cite{BESIII:nkspi} and the Cabibbo-suppressed (CS) decay $\Lambda_{c}^{+}\rightarrow n\pi^{+}$~\cite{BESIII:npi}. These studies provide critical tests for isospin and SU(3) symmetries in charmed-baryon decays, the violation of which could lead to enhanced  $C\!P$-violation effects~\cite{Jia:2019zxi,Saur:2020rgd}.
Taking the BFs of $\Lambda^+_c\to nK^0_S\pi^+$,  $\Lambda^+_c\to pK^0_S\pi^0$ and $\Lambda^+_c\to p K^-\pi^+$ from the Particle Data Group (PDG)~\cite{pdg2021}, the amplitudes of these three decays satisfy the triangle relation and obey isospin symmetry~\cite{Wang2016}.
However, the ratio of the BF of $\Lambda_{c}^{+}\rightarrow n\pi^{+}$ to that of  $\Lambda_{c}^{+}\rightarrow p\pi^{0}$ is measured to be larger than 7.2 which disagrees with the SU(3) flavor symmetry model prediction~\cite{Wang2016, Geng2018_2, Sharma:1996sc}.
Precise measurements of the BFs of additional  $\Lambda_c^+$ decays to  final states involving a neutron, such as $ \Lambda_c^+\to n \pi^+\pi^0$, $\Lambda_c^+\to n\pi^{+}\pi^{-}\pi^{+}$, and $\Lambda_c^+\to nK^{-}\pi^{+}\pi^{+}$, are highly desirable to allow for futher tests of isospin and SU(3) symmetries in charmed-baryon decays.
Throughout this paper, charge-conjugate modes are implicitly assumed.

In this paper, two CS decays $\Lambda_{c}^{+}\to n\pi^{+}\pi^{0}$, $\Lambda_{c}^{+}\to n\pi^{+}\pi^{-}\pi^{+}$, and the CF decay $\Lambda_{c}^{+}\to nK^{-}\pi^{+}\pi^{+}$ are studied by employing the double-tag (DT) method \cite{MARK-III:1989dea}, where the neutron is reconstructed via the missing-mass technique.
Our analysis is based on electron-positron annihilation data samples collected at seven center-of-mass (c.m.) energies from $\sqrt{s}=4599.53~\mev$ to $4698.82~\mev$ by the BESIII detector. These data samples correspond to an integrated luminosity of 4.5 fb$^{-1}$ \cite{BESIII:energy0,BESIII:energy1,BESIII:energy2,BESIII:energy3}, as listed in Table~\ref{tab:luminosity}.
Within this energy range, the electron-positron collisions provide a clean environment for the  production $\Lambda_c^+\bar{\Lambda}_c^{-}$ pairs, which offers a unique opportunity to carry out model-independent measurements of the BFs of various $\Lambda_{c}^{+}$ decays involving neutrons. The DT method allows us to measure the BFs without any theoretical input or external information on the cross section of $\Lambda_{c}^{+}$ production.
	\begin{center}
		\footnotesize
		\tabcaption{The c.m. energies and integrated  luminosities for the data samples.}
		\begin{tabular}{c|r@{.}l}
			\hline\hline
			$\sqrt{s}$ (MeV)    &  \multicolumn{2}{c}{Int. luminosity (pb$^{-1}$)} \\ \hline
			4599.53		        &   $586.90\pm0$&$10\pm3.90$  \\
			4611.86		        &   $103.65\pm0$&$05\pm0.55$  \\
			4628.00		        &   $521.53\pm0$&$11\pm2.76$  \\
			4640.91		        &   $551.65\pm0$&$12\pm2.92$  \\
			4661.24		        &   $529.43\pm0$&$12\pm2.81$  \\
			4681.92		        &   $1667.39\pm0$&$21\pm8.84$ \\
			4698.82		        &   $535.54\pm0$&$12\pm2.84$  \\
			\hline \hline
		\end{tabular}
		\label{tab:luminosity}
	\end{center}

The $\bar{\Lambda}^{-}_{c}$ baryons are reconstructed with twelve exclusive hadronic decay modes, as listed in Table~\ref{tab:4680}. This data set is referred to as the single-tag (ST) sample. 
The $\pi^0$, $K_{S}^{0}$, $\bar{\Lambda}$, $\bar{\Sigma}^0$, and $\bar{\Sigma}^-$ particles are reconstructed via individual dominant decay modes.
Those events in which any of the signal decays $\Lambda_c^+\to n \pi^+\pi^0$, $\Lambda_{c}^{+}\to n\pi^{+}\pi^{-}\pi^{+}$, and $\Lambda_{c}^{+}\to nK^{-}\pi^{+}\pi^{+}$ are reconstructed in the system recoiling against the ST $\bar{\Lambda}_{c}^{-}$ candidates are denoted as  DT candidates.
\begin{table*}[!htp]
	\begin{center}
		\footnotesize
		\caption{Requirements on $\Delta{}E$, ST yields, ST and DT efficiencies  for the signal decays of $\Lambda_c^+\to n \pi^+\pi^0$ ($\mathcal{E}$), $\Lambda_c^+\to n\pi^{+}\pi^{-}\pi^{+}$ ($\mathcal{D}$) and $\Lambda_c^+\to nK^{-}\pi^{+}\pi^{+}$ ($\mathcal{F}$) at $\sqrt{s}=4681.92~\mev$. The uncertainties are statistical only. The quoted efficiencies do not include any subdecay BFs. Entries of ``$\cdots$" are for the cases where knowledge of the DT efficiencies is not required in the analysis.}
			\begin{tabular}{l|c|r@{$\pm$}l|c|c|c|c}
				\hline \hline
				Tag mode  & $\Delta{}E$($~\mev$) & \multicolumn{2}{c|}{$N_{i}^{\rm ST}$} & $\varepsilon_{i}^{\rm ST} (\%)$ & $\varepsilon_{i}^{\rm DT}(\mathcal{E}) (\%)$ & $\varepsilon_{i}^{\rm DT}(\mathcal{D}) (\%)$ &   $\varepsilon_{i}^{\rm DT}(\mathcal{F}) (\%)$ \\ \hline
				\textbf{$\Modea$}   & $(-21,18)$  &  3376&61    &  $48.6$ & $12.58$  &  $11.55$  &  $18.50$   \\
				\textbf{$\Modeb$}   & $(-29,26)$  &  17508&147  &  $47.1$ & $10.77$  &  $10.59$  &  $16.61$   \\
				\textbf{$\Modec$}   & $(-49,34)$  &  1785&63    &  $19.7$ & $4.74 $  &   $ 4.65$ &  $ 7.15$   \\
				\textbf{$\Moded$}   & $(-34,31)$  &  1511&57    &  $20.7$ & $ 4.22$  &  $ 4.13$  &  $ 5.99$   \\
				\textbf{$\Modee$}   & $(-60,41)$  &  5111&128   &  $17.8$ & $\cdots$ &  $ 4.13$  &  $ 6.60$   \\
				\textbf{$\Modeaa$}  & $(-23,21)$  &  2074&48    &  $41.3$ & $ 9.95$  &  $ 9.12$  &  $15.08$   \\
				\textbf{$\Modebb$}  & $(-50,41)$  &  4380&88    &  $18.2$ & $ 4.05$  &  $ 3.85$  &  $ 6.00$   \\
				\textbf{$\Modedd$}  & $(-40,36)$  &  2059&61    &  $13.9$ & $ 2.80$  &  $ 2.67$  &  $ 3.93$   \\
				\textbf{$\Modeaaa$} & $(-33,31)$  &  1398&42    &  $25.9$ & $ 6.31$  &  $ 5.89$  &  $ 8.46$   \\
				\textbf{$\Modeccc$} & $(-67,32)$  &  879&43     &  $22.5$ & $ 5.78$  &  $ 5.21$  &  $ 8.22$   \\
				\textbf{$\Modeddd$} & $(-40,32)$  &  3027&88    &  $22.1$ & $ 5.40$  &  $ 5.21$  &  $ 7.77$   \\
				\textbf{$\Modef$}   & $(-26,20)$  &  1596&80    &  $53.6$ & $\cdots$ &  $\cdots$ &  $17.35$   \\
				\hline  \hline
			\end{tabular}
		\label{tab:4680}
	\end{center}
\end{table*}

\section{BESIII detector and Monte Carlo simulation}
The BESIII detector~\cite{BESIII:2009fln} records symmetric $e^+e^-$ collisions 
provided by the BEPCII storage ring~\cite{Yu:2016cof}, which operates in the center-of-mass energy range from 2.0~$\gev$ to 4.95~$\gev$.
BESIII has collected large data samples in this energy region~\cite{BESIII:2020nme}. The cylindrical core of the BESIII detector covers 93\% of the full solid angle and consists of a helium-based multilayer drift chamber~(MDC), a plastic scintillator time-of-flight
system~(TOF), and a CsI(Tl) electromagnetic calorimeter~(EMC),
which are all enclosed in a superconducting solenoidal magnet
providing a 1.0~T magnetic field. The solenoid is supported by an
octagonal flux-return yoke with resistive plate counter muon
identification modules interleaved with steel. 
The charged-particle momentum resolution at $1~\gevc$ is
$0.5\%$, and the specific ionization energy loss (d$E$/d$x$) resolution is $6\%$ for electrons
from Bhabha scattering. The EMC measures photon energies with a
resolution of $2.5\%$ ($5\%$) at $1~\gev$ in the barrel (end-cap)
region. The time resolution in the TOF barrel region is 68~ps, while
that in the end-cap region is 60~ps~\cite{etof1,etof2,etof3}. High-statistics Monte Carlo (MC) simulation samples for the process $e^{+}e^{-}\to {\rm inclusive}$ are produced with the KKMC generator\cite{Jadach:2000ir} by incorporating the initial-state radiation (ISR) effects and the beam-energy spread. 
The inclusive MC simulation sample, which consists of $\Lambda_c^{+}\bar{\Lambda}_c^{-}$ events, $D_{(s)}^{(*)}$ production, ISR return to lower-mass $\psi$ states, and continuum processes $e^{+}e^{-}\rightarrow q\bar{q}$ ($q=u,d,s$), is generated to determine the ST detection efficiencies and estimate the potential background.
For this MC simulation sample, all the known decay modes of charmed hadrons and charmonia are modeled with {\sc evtgen}~\cite{Lange:2001uf, Ping:2008zz} using BFs taken from the PDG~\cite{pdg2021}, and the remaining unknown decays are modeled with {\sc lundcharm}~\cite{Chen:2000tv}. 
Furthermore, exclusive DT signal MC simulation events, where the $\bar{\Lambda}_{c}^{-}$ decays into any of the tag modes and the $\Lambda_{c}^{+}$ decays into any of the signal modes of $n\pi^{+}\pi^{0}$, $n\pi^{+}\pi^{-}\pi^{+}$ or $nK^{-}\pi^{+}\pi^{+}$, are used to determine the DT dectection efficiencies.
The Born cross sections are taken into account when producing the MC simulation sample of $\Lambda_c^{+}\bar{\Lambda}_c^{-}$ pairs. 
The $\Lambda_{c}^{+}\to n\pi^{+}\pi^{0}$ and $\Lambda_{c}^{+}\to n\pi^{+}\pi^{-}\pi^{+}$ signal MC samples are simulated evenly distributed in phase space since the angular distributions, momentum distributions and the two-body invariant mass distributions of the final state particles of the signal MC simulation samples are in good agreement with data with the current size of data set. For the signal MC simulation sample of $\Lambda^+_c\to nK^{-}\pi^{+}\pi^{+}$, the key kinematic distributions mentioned above have been reweighted to agree with those of data.
All final tracks and photons are fed into a GEANT4-based~\cite{geant4_1,geant4_2} detector simulation package, which includes the geometric description of the BESIII detector.

\section{Analysis}

Charged tracks detected in the MDC must satisfy $|\!\cos\theta|<0.93$ (where $\theta$ is defined with respect to the $z$-axis, which is the symmetry axis of the MDC) and have a distance of closest approach to the interaction point (IP) of less than 10 cm along the beam axis and less than 1 cm in the perpendicular plane, except for those used for reconstructing $K_{S}^{0}$ and $\bar{\Lambda}$ decays.
Particle identification (PID) for charged tracks combines measurements of the energy deposited in the MDC (d$E/$d$x$) and the flight time in the TOF to form likelihoods $\mathcal{L} (h) (h = p, K, \pi)$ for each hadron $h$ hypothesis.
Tracks are identified as protons when their likelihoods satisfy $\mathcal{L}(p)>\mathcal{L}(K)$ and $\mathcal{L}(p)>\mathcal{L}(\pi)$, while charged kaons and pions are identified by comparing the likelihoods for the kaon and pion hypotheses, $\mathcal{L}(K)>\mathcal{L}(\pi)$ or $\mathcal{L}(\pi)>\mathcal{L}(K)$, respectively.

Neutral showers are reconstructed in the EMC.
Showers not associated with any charged track are identified as photon candidates. The deposited energy of each shower in the EMC must be greater than $25~\mev$ in the barrel region, corresponding to the polar angle $|\!\cos\theta|<0.80$, and greater than $50~\mev$ in the end-cap region, corresponding to $0.86<|\!\cos\theta|<0.92$. The EMC time difference from the event start time is required to be less than 700~ns, to suppress electronic noise and showers unrelated to the event.
The $\pi^{0}$ candidates are reconstructed from photon pairs with invariant masses within $115~\mevcc<M(\gamma\gamma)<150~\mevcc$.
To improve momentum resolution, a kinematic fit constraining the photon pairs to the $\pi^0$ known mass is performed and the resulting four-momentum of the $\pi^0$ candidate is used for further analysis.

Candidates for $K_{S}^{0}$ and $\bar{\Lambda}$ decays are formed by $\pi^{+}\pi^{-}$ and $\bar{p}\pi^{+}$ combinations, respectively. 
For these tracks, their distances of closest approaches to the IP must be within $\pm$20\,cm along the beam direction.
No distance constraint in the transverse plane is required.
The charged pion is not subjected to the PID requirement described above, while the proton PID is imposed.
The two final-state tracks are constrained to originate from a common decay vertex by requiring the $\chi^2$ of the vertex fit to be less than 100.
Furthermore, the decay vertex is required to be separated from the IP by a distance of at least twice the fitted vertex resolution.
The fitted momenta of the $\pi^{+}\pi^{-}$ and $\bar{p}\pi^{+}$ combinations are used in the subsequent analysis.
We require $487~\mevcc<M(\pi^+\pi^{-})<511~\mevcc$ and $1111~\mevcc<M(\bar{p}\pi^{+})<1121\,~\mevcc$ to select $K_{S}^{0}$ and $\bar{\Lambda}$ candidates, respectively, which corresponds to a window of about three times the standard deviations either side of the known masses.
The $\bar{\Sigma}^0$ and $\bar{\Sigma}^-$ candidates are reconstructed with the $\gamma\bar{\Lambda}$ and $\bar{p}\pi^{0}$ combinations with invariant masses being in the intervals of $1179~\mevcc<M(\gamma\bar{\Lambda})<1203~\mevcc$ and $1176~\mevcc<M(\bar{p}\pi^{0})<1200~\mevcc$, respectively.

For the tag modes $\bar \Lambda^-_c\to $ $\Modec$, $\bar \Lambda^-_c\to \Moded$, $\bar \Lambda^-_c\to \Modeddd$, and $\bar \Lambda^-_c\to $ $\Modef$, possible backgrounds with $\bar{\Lambda}\to \bar{p}\pip$ are rejected by requiring $M(\bar{p}\pip)$ to be outside the range $(1110, 1120)~\mevcc$.
In the selection of   $\bar \Lambda^-_c\to \Modec$ decays, candidate events within the range $1170~\mevcc<M(\bar{p}\pi^0)<1200~\mevcc$ are excluded to suppress background from  $\bar{\Sigma}^-$ decays.
To remove $K_{S}^{0}$ decays in the selection of  $\bar \Lambda^-_c\to \Modedd$, $\bar \Lambda^-_c\to \Modeccc$, $\bar \Lambda^-_c\to \Modeddd$, and $\bar \Lambda^-_c\to $ $\Modef$ candidates, the invariant masses of any  $\pip\pim$ and $\pi^0\pi^0$ pairs are required to lie outside of the range $(480, 520)~\mevcc$.
 MC simulation studies indicate that peaking backgrounds and cross feeds among the twelve ST modes are negligible after applying the above veto procedures.

The ST $\bar{\Lambda}_{c}^{-}$ baryons are identified using the beam-constrained mass $M_{\rm BC}\equiv \sqrt{E_{\rm beam}^2/c^4-p^2/c^2}$, where  $E_{\rm beam}$ is the average value of the $e^+$ and $e^-$ beam energies and $p$ is the measured $\bar{\Lambda}_{c}^{-}$ momentum in the c.m. system of the $\ee$ collision.
To improve the signal purity, the energy difference $\Delta{}E \equiv E - E_{\rm beam}$ for the $\bar{\Lambda}_{c}^{-}$ candidate is required to fulfil a mode-dependent $\Delta{}E$ requirement shown in Table~\ref{tab:4680}, corresponding to approximately three times the resolutions.
Here, $E$ is the total reconstructed energy of the $\bar{\Lambda}_{c}^{-}$ candidate.
For each ST mode, if more than one candidate satisfies the above requirements, we select the one with the minimal $|\Delta{}E|$.
Figure~\ref{fig:ST_datafit_4680} shows the $M_{\rm BC}$ distributions of various ST modes for the data sample at $\sqrt{s}=4681.92~\mev$, where a clear $\bar{\Lambda}_{c}^{-}$ signal peak can be seen in each mode.

To obtain the ST yields, unbinned maximum-likelihood fits  are performed to the $M_{\rm BC}$ distributions, where the signal shapes are modeled with the MC-simulated shape convolved with a Gaussian function representing the resolution difference between data and MC simulation, and the background shapes are described by an ARGUS function~\cite{ARGUS:1990hfq}.
The fit results for the data sample at $\sqrt s=$4681.92~MeV are shown in Fig.~\ref{fig:ST_datafit_4680}.
The fits to the $M_{\rm BC}$ distributions for the other six data samples at different c.m.~energies are shown in Appendix B..
Candidates with $M_{\rm BC}\in(2275, 2300)~\mevcc$ for the data sample at $\sqrt{s}=4599.53~\mev$, $M_{\rm BC}\in(2275, 2306)~\mevcc$ for the data samples at $\sqrt{s}=4611.86~\mev, 4628.00~\mev, 4640.91~\mev$, and $M_{\rm BC}\in(2275, 2310)~\mevcc$ for the data samples at $\sqrt{s}=4661.24~\mev, 4681.92~\mev, 4698.82~\mev$ are retained for further analysis. 
The differences in selection requirements between data sets are necessary as the resolution and effects of ISR vary with collision energy.
The fitted ST yields, ST and DT efficiencies for each ST mode at $\sqrt s=4681.92$ MeV are summarized in Table~\ref{tab:4680}; those for the other c.m. energy points can be found in Appendix B.
\begin{center}
	\centering
	\includegraphics[width=0.45\textwidth]{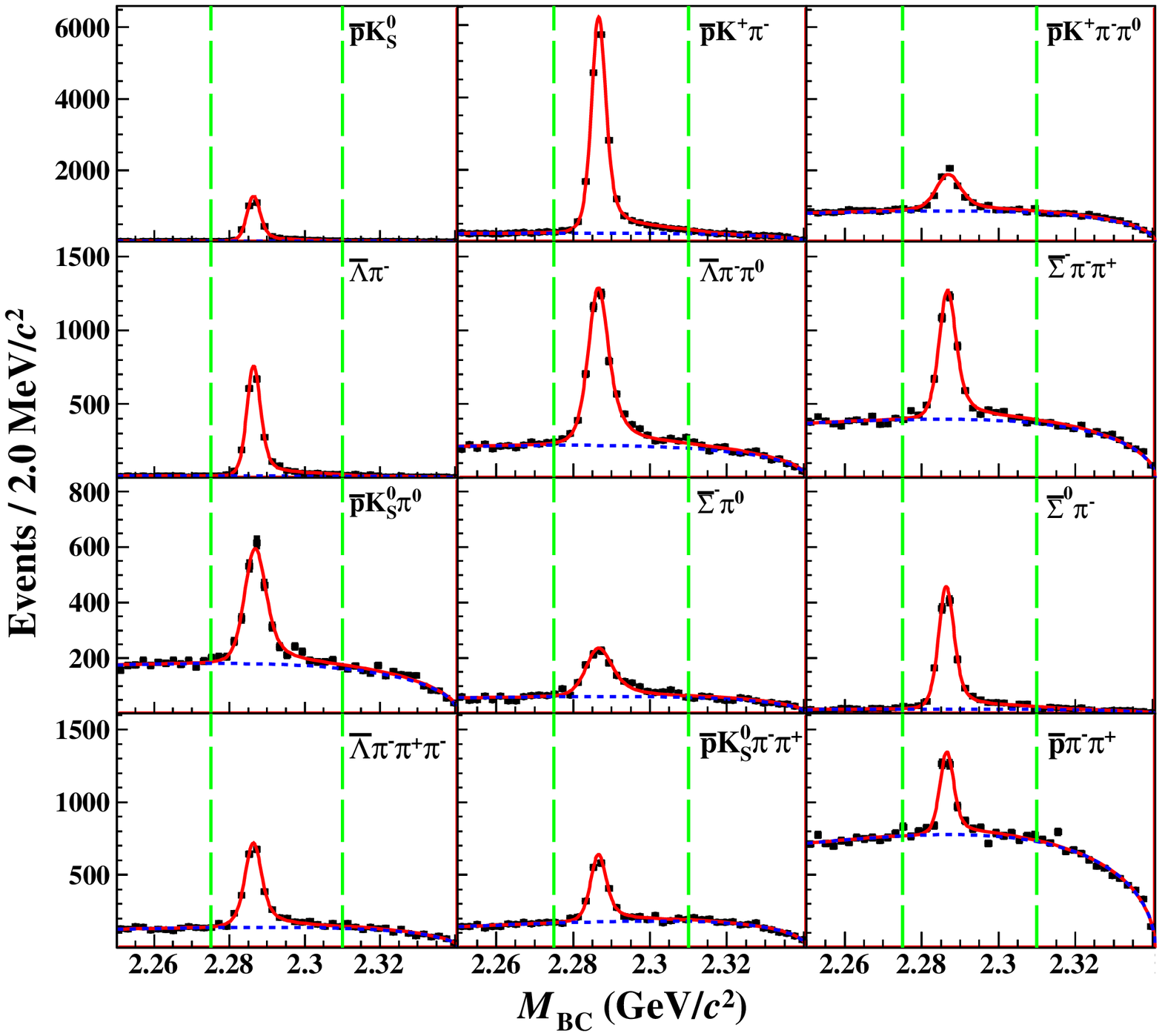}
	\figcaption{ The $M_{\rm BC}$ distributions of various ST modes for the data sample at $\sqrt{s}$=4681.92$~\mev$.  The points with error bars represent data. The (red) solid curves indicate the fit results and the (blue) dashed curves describe the fitted background shapes. The ranges between green dashed lines are the signal regions.} 
	\label{fig:ST_datafit_4680}
\end{center}

Searches are performed for the decays $\Lambda_c^+\to n\pi^{+}\pi^{0}$, $\Lambda_c^+\to n\pi^{+}\pi^{-}\pi^{+}$ and $\Lambda_c^+\to nK^{-}\pi^{+}\pi^{+}$ among the remaining tracks and showers recoiling against the ST $\bar{\Lambda}_{c}^{-}$ candidates.
In the case of $\Lambda_{c}^{+}\to n\pi^{+}\pi^{0}$, the event is allowed to contain only one pion with opposite charge to the tagged $\bar{\Lambda}_{c}^{-}$ satisfying the same selection criteria as described above.
 The $\pi^0$ candidate giving rise to the smallest $\chi^{2}$ for the mass-constrained kinematic fit is retained.
When searching for  $\Lambda_{c}^{+}\to n\pi^{+}\pi^{-}\pi^{+}$ and $\Lambda_{c}^{+}\to n K^{-}\pi^{+}\pi^{+}$ decays, events are selected with only three remaining charged tracks, satisfying the desired charge and PID criteria.
In each of the decays, the kinematic variable $M_{\rm miss} \equiv \sqrt{E_{\rm miss}^{2}/c^4-|\vec{{p}}_{\rm miss}|^{2}/c^2}$ is used to infer the presence of the undetected neutron.
Here, $E_{\rm miss}$ and $\vec{{p}}_{\rm miss}$ are calculated by $E_{\rm miss} \equiv E_{\rm beam}-E_{\rm rec}$ and $\vec{{p}}_{\rm miss} \equiv \vec{{p}}_{\Lambda_{c}^{+}} - \vec{{p}}_{\rm rec}$, where $E_{\rm rec}$($\vec{{p}}_{\rm rec}$) is the energy (momentum) of the reconstructed final-state particles in the $\ee$ c.m. system.
The momentum of the $\Lambda_{c}^{+}$ baryon $\vec{{p}}_{\Lambda_{c}^{+}}$ is calculated by $\vec{{p}}_{\Lambda_{c}^{+}} \equiv -\hat{p}_{\rm tag} \sqrt{E_{\rm beam}^{2}/c^2-m_{\Lambda_{c}^{+}}^{2} c^2}$, where $\hat{p}_{\rm tag}$ is the momentum direction of the ST $\bar{\Lambda}_{c}^{-}$ and $m_{\Lambda_{c}^{+}}$ is the known mass of the $\Lambda_{c}^{+}$ baryon~\cite{pdg2021}.
In the case of  $\Lambda_{c}^{+}\to n\pi^{+}\pi^{0}$, $\Lambda_{c}^{+}\to n\pi^{+}\pi^{-}\pi^{+}$, and $\Lambda_{c}^{+}\to n K^{-}\pi^{+}\pi^{+}$ decays, the $M_{\rm miss}(\pi^{+}\pi^{0})$, $M_{\rm miss}(\pi^{+}\pi^{-}\pi^{+})$ and $M_{\rm miss}(K^{-}\pi^{+}\pi^{+})$ spectra are expected to peak around the known mass of the neutron $\ie$ at $939.6~\mevcc$~\cite{pdg2021}.
A study of the inclusive MC simulation sample reveals that the dominant background events for the signal mode $\Lambda_{c}^{+}\to n\pi^{+}\pi^{0}$ are from the processes $\Lambda_{c}^{+}\to\Lambda\pi^{+}$ with $\Lambda\to n\pi^{0}$, $\Lambda_{c}^{+}\to\Sigma^{+}\pi^{0}$ with $\Sigma^{+}\to n\pi^{+}$ and $\Lambda_{c}^{+}\to \Sigma^{0}\pi^{+}$ with $\Sigma^{0}\to\gamma\Lambda(\to n\pi^{0})$.
In order to reject these background events for $\Lambda_{c}^{+}\to n\pi^{+}\pi^{0}$, the following selection criteria are applied: $M_{\rm miss}(\pi^{+})>$ 1300 $~\mevcc$ and $M_{\rm miss}(\pi^{0})>$1370$~\mevcc$.
For the signal mode $\Lambda_{c}^{+}\to n\pi^{+}\pi^{-}\pi^{+}$, the peaking backgrounds from the decays $\Lambda_{c}^{+}\to nK_{S}^{0}(\to\pi^{+}\pi^{-})\pi^{+}$, $\Lambda_{c}^{+}\to\Sigma^{+}(\to n\pi^{+})\pi^{+}\pi^{-}$, and $\Lambda_{c}^{+}\to \Sigma^{-}(\to n\pi^{-})\pi^{+}\pi^{+}$ are suppressed 
by requiring $M_{\pi^{+}\pi^{-}}\notin (487, 511)~\mevcc$, $M_{\rm miss}(\pi^{+}\pi^{-})\notin(1150, 1250)~\mevcc$ and $M_{\rm miss}(\pi^{+}\pi^{+})\notin(1150, 1250)~\mevcc$.
Note that, based on the study of the inclusive MC simulation sample, ten ST modes (after excluding $\bar{\Lambda}_{c}^{-}\to\Modee$ and $\bar{\Lambda}_{c}^{-}\to\Modef$) are used in the analysis of the decay $\Lambda_c^+\to n \pi^+\pi^0$, and eleven ST modes (after excluding $\bar{\Lambda}_{c}^{-}\to\Modef$) are used in the analysis of the decay $\Lambda_c^+\to n\pi^{+}\pi^{-}\pi^{+}$ to improve the background level in the $M_{\rm miss}$ spectra.

\begin{figure*}[!htb]
\begin{center}
\centering
\includegraphics[width=0.45\textwidth]{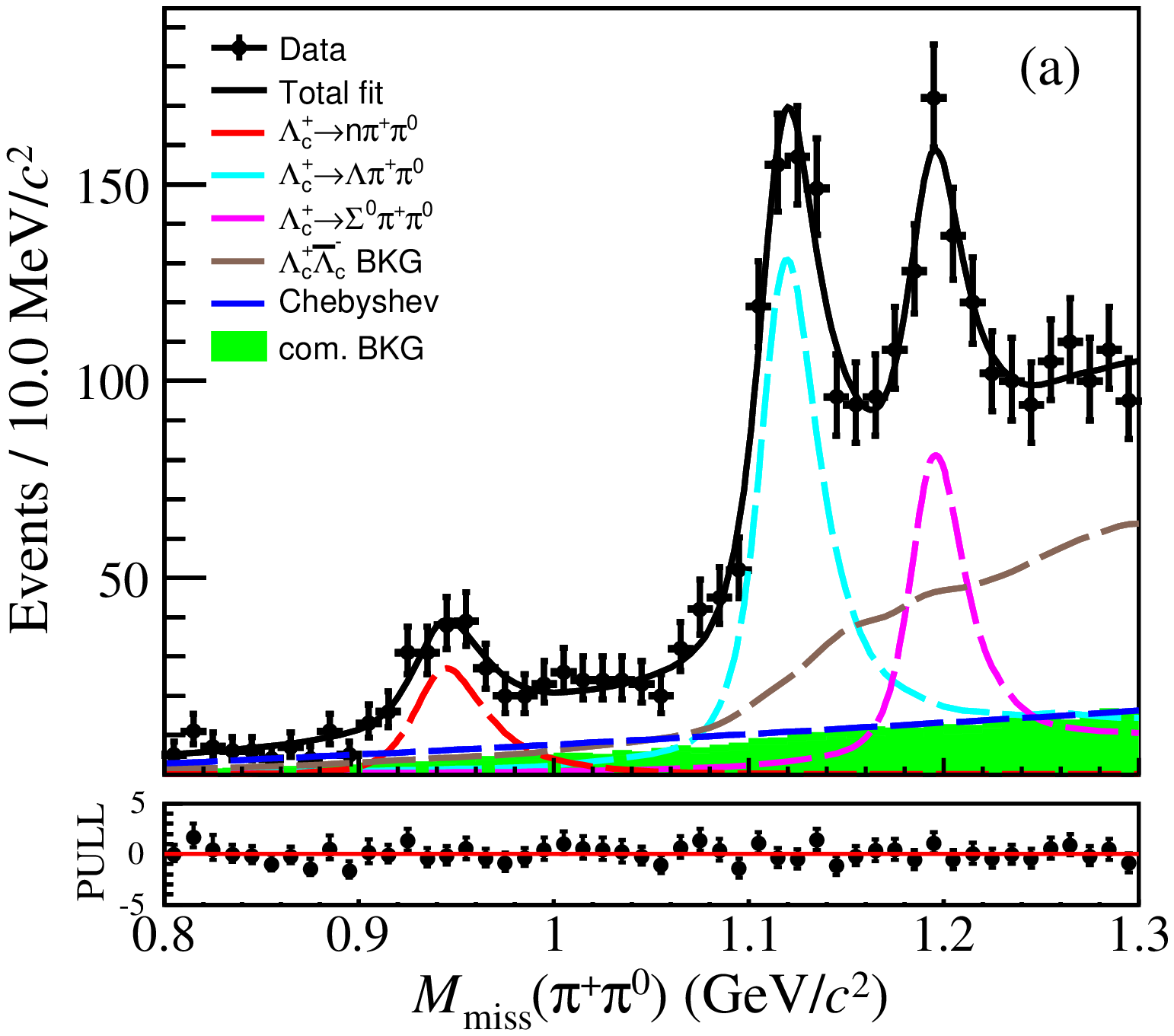}
\includegraphics[width=0.45\textwidth]{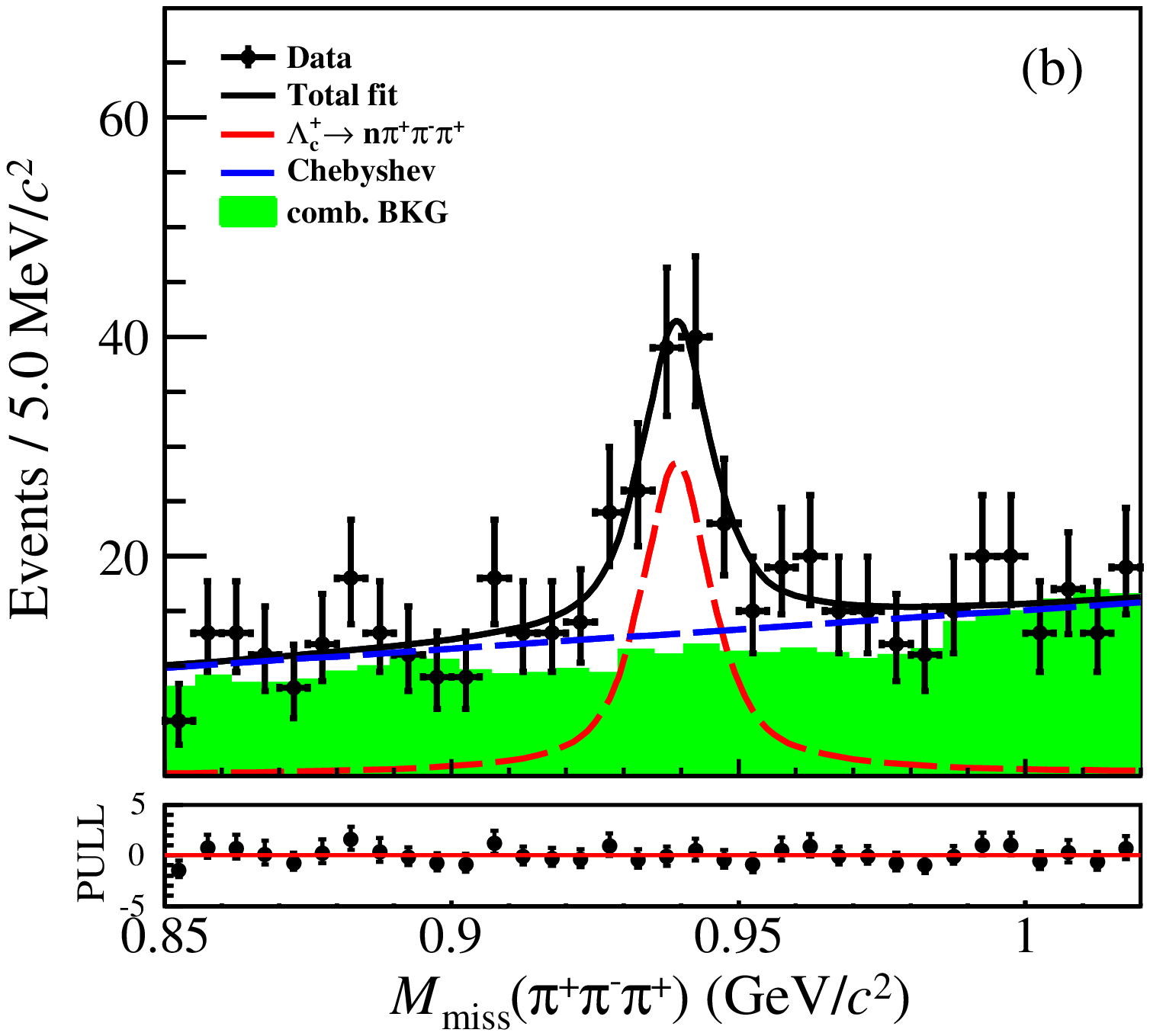}
\includegraphics[width=0.45\textwidth]{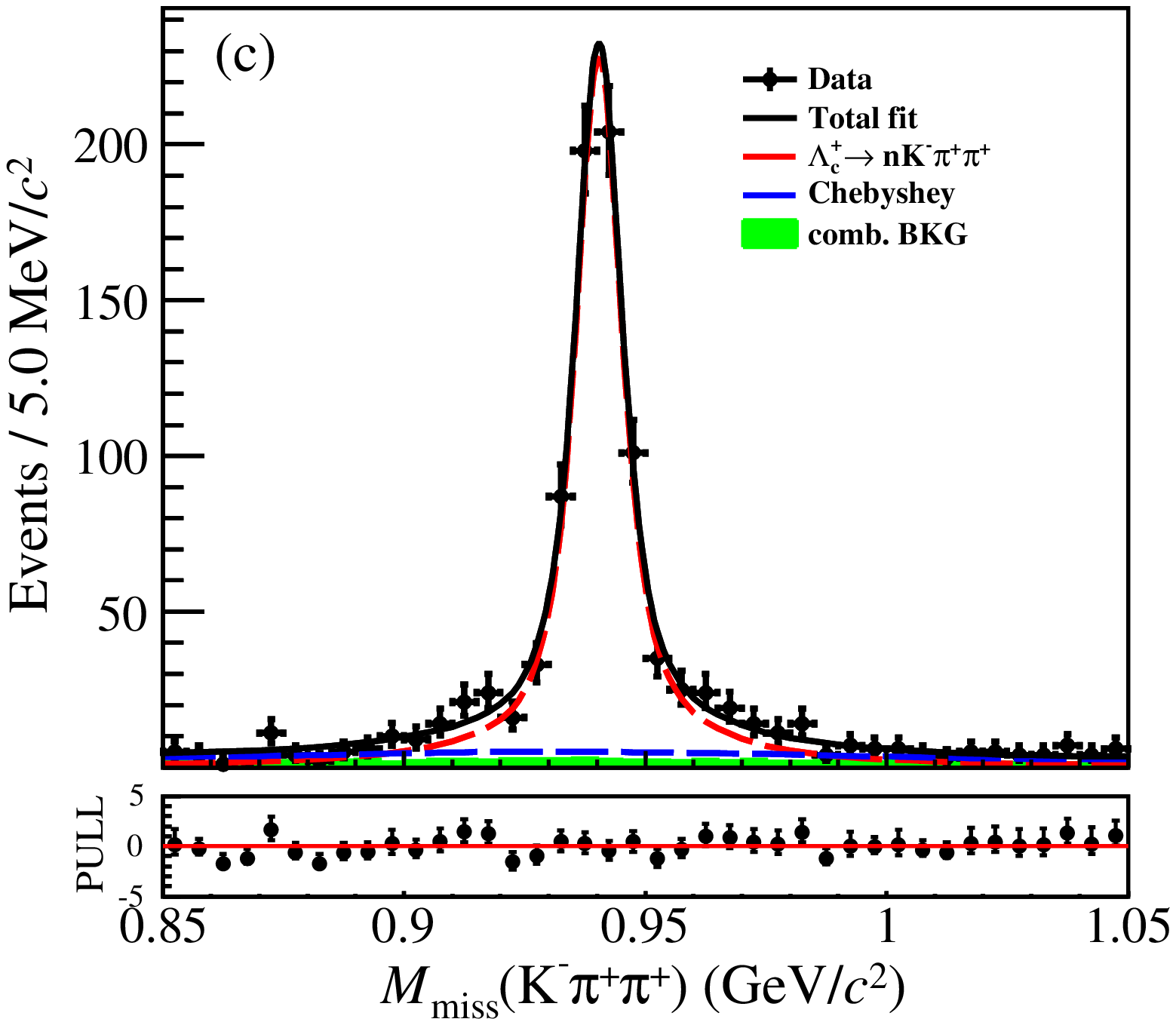}
\figcaption{The $M_{\rm miss}$ distributions of the surviving DT candidate events for (a) $\Lambda_{c}^{+}\to n\pi^{+}\pi^{0}$, (b) $\Lambda_c^+\to n\pi^{+}\pi^{-}\pi^{+}$ and (c) $\Lambda_{c}^{+}\to nK^{-}\pi^{+}\pi^{+}$ decays with fit results overlaid.
	The points with error bars are data combined from seven c.m. energy points. The solid black curves are the fit results.
	The red, cyan, and pink dashed curves indicate the neutron, $\Lambda$, and $\Sigma$ signal shapes, respectively.
	The brown dashed curve is the $\Lambda_{c}^{+}\bar{\Lambda}_{c}^{-}$ background shape for $\Lambda_{c}^{+}\to n\pi^{+}\pi^{0}$.
	The blue dashed curves represent the fitted combinatorial background shape.
	The green histograms are the simulated combinatorial background shapes from the inclusive MC sample. 
	The plots in the bottom of each graphs show the pull value of each bin, in which the values are expected to fluctuate around 0.
}
\label{fig:DT_datafit}
\end{center}
\end{figure*}

The resulting $M_{\rm miss}$ distributions of the DT candidate events summed over all data samples at seven c.m. energies are shown in Fig.~\ref{fig:DT_datafit}.
A peak around the neutron mass is observed in Fig.~\ref{fig:DT_datafit}(a) representing the $\Lambda_{c}^{+}\to n\pi^{+}\pi^{0}$ signal.
Moreover, there are two prominent structures peaking around the $\Lambda$ and $\Sigma^{0}$ mass regions, which correspond to the CF decays $\Lambda_{c}^{+}\to \Lambda\pi^{+}\pi^{0}$ and $\Lambda_{c}^{+}\to \Sigma^{0}\pi^{+}\pi^{0}$, respectively.
 Significant signal peaks around the neutron mass are also observed for $\Lambda_{c}^{+}\to n\pi^{+}\pi^{-}\pi^{+}$ and $\Lambda_{c}^{+}\to nK^{-}\pi^{+}\pi^{+}$ in Fig.~\ref{fig:DT_datafit}(b) and Fig.~\ref{fig:DT_datafit}(c), respectively.

The total DT signal yield is obtained by performing an unbinned maximum-likelihood fit on the $M_{\rm miss}$ distribution. 
The neutron, $\Lambda$ and $\Sigma^0$ signals are modeled by individual MC-derived shapes convolved with Gaussian functions that account for the shift and resolution difference between data and MC simulation.
The Gaussian-function parameters are left free and are shared by the three signal processes in the fit to the $M_{\rm miss}(\pi^{+}\pi^{0})$ spectra.
The potential background events in the $M_{\rm miss}$ distributions are classified into two categories. Those directly originating from continuum hadron production in the $\ee$ annihilation are denoted as $q\bar{q}$ background. Those from $\ee\to\Lambda_c^+\bar{\Lambda}_{c}^{-}$ events excluding the contributions from the corresponding signal process are referred to as $\Lambda_c^+\bar{\Lambda}_{c}^{-}$ background.
In the fit to the $M_{\rm miss}(\pi^{+}\pi^{0})$ distribution, the $q\bar{q}$ background is described by a second-order Chebyshev polynomial with free parameters and the $\Lambda_c^+\bar{\Lambda}_{c}^{-}$ backgroud shape is taken from the inclusive MC simulation sample.
For the fit to the $M_{\rm miss}(\pi^{+}\pi^{-}\pi^{+})$ spectrum, only one first-order Chebyshev polynomial with free parameters is used to model all the background.
Similarly, for the $M_{\rm miss}(K^{-}\pi^{+}\pi^{+})$ distribution, one second-order Chebyshev function with fixed parameters is taken as the background shape where the parameters are derived from the fit to the inclusive MC simulation sample.
The DT signal and background yields are left free in the fit to the $M_{\rm miss}(\pi^{+}\pi^{0})$, $M_{\rm miss}(\pi^{+}\pi^{-}\pi^{+})$ and $M_{\rm miss}(K^{-}\pi^{+}\pi^{+})$ spectra.
Figure~\ref{fig:DT_datafit} shows the results of the fits to the $M_{\rm miss}$ distributions.
From these fits, we determine the DT signal yields of  $\Lambda_{c}^{+}\to n\pi^{+}\pi^{0}$, $\Lambda_{c}^{+}\to n\pi^{+}\pi^{-}\pi^{+}$, and $\Lambda_{c}^{+}\to nK^{-}\pi^{+}\pi^{+}$ to be $N^{\rm DT}_{n\pi^{+}\pi^{0}}=150.9\pm21.4$, $N^{\rm DT}_{n\pi^{+}\pi^{-}\pi^{+}}=120.6\pm17.9$ and $N^{\rm DT}_{nK^{-}\pi^{+}\pi^{+}}=805.8\pm33.1$, respectively, where the uncertainties are statistical only. 
The statistical significances of the $\Lambda_{c}^{+}\to n\pi^{+}\pi^{0}$, $\Lambda_{c}^{+}\to n\pi^{+}\pi^{-}\pi^{+}$ and $\Lambda^+_c\to nK^{-}\pi^{+}\pi^{+}$ signals are $7.9\sigma$, $7.8\sigma$, and $>10\sigma$, respectively, 
which are evaluated by the changes in the likelihoods between the nominal fit and the fit with the signal yield set to zero, and accounting for the change in the number of degrees of freedom.

The BFs of the decays $\Lambda_{c}^{+}\to n\pi^{+}\pi^{0}$, $\Lambda_{c}^{+}\to n\pi^{+}\pi^{-}\pi^{+}$, and $\Lambda_{c}^{+}\to nK^{-}\pi^{+}\pi^{+}$ are determined as 
\begin{equation}
	\label{eq:br}
	\mathcal{B}=\frac{N^{\rm DT}}{\sum_{ij} N_{ij}^{\mathrm{ST}}\cdot (\epsilon_{ij}^{\mathrm{DT}}/\epsilon_{ij}^{\mathrm{ST}})\cdot\mathcal{B}_{\rm int} },  
\end{equation}
where $i$ and $j$ represent the ST modes and the data samples at different c.m.~energies, respectively.
The factor  $\mathcal{B}_{\rm int}$ is $(98.823\pm0.034)\%$, which is the BF of $\pi^{0}\to\gamma\gamma$ \cite{pdg2021}, is only present for $\Lambda_{c}^{+}\to n\pi^{+}\pi^{0}$.
$N_{ij}^{\mathrm{ST}}$, $\epsilon_{ij}^{\mathrm{ST}}$, and $\epsilon_{ij}^{\mathrm{DT}}$ are the ST yields, ST efficiencies, and DT efficiencies, respectively.
The detection efficiencies $\epsilon_{ij}^{\mathrm{ST}}$ and $\epsilon_{ij}^{\mathrm{DT}}$ are estimated from the inclusive MC simulation sample and exclusive DT signal MC simulation samples, respectively. 
The ST and DT efficiencies for the data sample at $\sqrt{s}=4681.92$~MeV are summarized in Table~\ref{tab:4680}.
The detection efficiencies for the other data samples are summarized in Appendix B.
The obtained BFs are summaried in Table~\ref{tab:bf}.

\begin{center}
		\scriptsize
		\tabcaption{The obtained BFs, where the first uncertainties are statistical and the second are systematic.}
		\begin{tabular}{l|c}
			\hline \hline
			Signal decay  &   $\mathcal{B}~(\%)$   \\
			\hline
			$\Lambda_{c}^{+}\rightarrow n\pi^{+}\pi^{0}$	 			& $0.64\pm0.09\pm0.02$ \\
			$\Lambda_{c}^{+}\rightarrow n\pi^{+}\pi^{-}\pi^{+}$         & $0.45\pm0.07\pm0.03$ \\
			$\Lambda_{c}^{+}\rightarrow nK^{-}\pi^{+}\pi^{+}$           & $1.90\pm0.08\pm0.09$ \\
			
			\hline \hline
		\end{tabular}
		\label{tab:bf}
\end{center}


 Most systematic uncertainties from the ST side cancel in the determination of the BFs, as is clear from Eqn.~\eqref{eq:br}. 
 However, effects from the signal side can lead to systematic bias, for example, the requirement of no extra charged track, tracking efficiency, PID efficiency, $\pi^{0}$ reconstruction, peaking background veto, $M_{\rm miss}$ fit, ST $\bar{\Lambda}_{c}^{-}$ yield, MC modeling, and MC sample size.
The systematic uncertainty due to the requirement of no extra charged track is assigned as 1.1\% from the study of a control sample of $\ee\to\Lambda_{c}^{+}\bar{\Lambda}_{c}^{-}$ with $\Lambda_{c}^{+}\to nK^{-}\pi^{+}\pi^{+}$ and $\bar{\Lambda}_{c}^{-}$ decays to tag modes.
The systematic uncertainties associated with the efficiencies of the tracking and PID of charged particles are estimated to be $1\%$ by using control samples of $\ee\to \pip\pip\pim\pim$ and $\ee\to K^+K^-\pi^+\pi^-$ events collected at c.m. energies above $\sqrt{s}=4.0\gev$.
The systematic uncertainty due to the $\pi^{0}$ reconstruction efficiency is assigned to be $1.0\%$ \cite{BESIII:hadron}.
The systematic uncertainty associated with the  BF of $\pi^{0}\to\gamma\gamma$ is $0.03\%$ \cite{pdg2021}, which is negligible.
In order to estimate the systematic uncertainties arising from the veto of peaking backgrounds involving $\Lambda$, $\Sigma^{+}$, $\Sigma^{0}$, $K_{S}^{0}$, and $\Sigma^{-}$, the corresponding resolutions in the MC simulation samples are corrected to agree with those in data, and the BFs are then re-evaluated with the updated MC simulation samples.
The deviations from the baseline BF measurements are taken as the associated systematic uncertainties for $\Lambda_{c}^{+}\to n\pi^{+}\pi^{0}$ and $\Lambda_{c}^{+}\to n\pi^{+}\pi^{-}\pi^{+}$, which are 1.6\% and 1.0\%, respectively.  
The systematic uncertainties from the fitted DT yields are 1.3\%, 2.2\%, and 0.9\% for $\Lambda_{c}^{+}\to n\pi^{+}\pi^{0}$, $n\pi^{+}\pi^{-}\pi^{+}$, and $nK^{-}\pi^{+}\pi^{+}$, respectively, which are estimated from varying the alternative polynomial descriptions for the $q\bar{q}$ and $\Lambda_{c}^{+}\bar{\Lambda}_{c}^{-}$ backgrounds, respectively. 
The systematic uncertainty in the total ST $\bar{\Lambda}_{c}^{-}$ yield is 0.1\% for $\Lambda_{c}^{+}\to n\pi^{+}\pi^{0}$, and 0.2\% for $\Lambda_{c}^{+}\to n\pi^{+}\pi^{-}\pi^{+}$, and $nK^{-}\pi^{+}\pi^{+}$. These uncertainties arise from the fluctuation of background together with a component coming from the fit to the $M_{\rm BC}$ distribution.
The systematic uncertainties arising from the MC modeling are investigated by reweighting the MC distribution to data, and they are assigned as the efficiency differences between the original and reweighted samples. 
The systematic uncertainties due to limited sample sizes of the MC samples are estimated to be 0.2\%.
Assuming that all the sources are uncorrelated, the total uncertainties are then taken to be 
the quadratic sums of the individual values, which are 3.1\%, 5.6\%, and 4.5\% for $\Lambda_{c}^{+}\to n\pi^{+}\pi^{0}$, $\Lambda_{c}^{+}\to n\pi^{+}\pi^{-}\pi^{+}$, and $\Lambda_{c}^{+}\to nK^{-}\pi^{+}\pi^{+}$, respectively.
All the above systematic uncertainties are summarized in Table~\ref{tab:sys_err}.
	\begin{center}
		\vspace{0.3cm}
		\tabcaption{Relative systematic uncertainties in BF measurements. ``$\cdots$" means the uncertainty is negligible. ``$-$" indicates cases where there is no uncertainty. }
		\resizebox{\linewidth}{!}{
		\begin{tabular}{l|c|c|c}
			\hline \hline
			Source            &  $n\pi^{+}\pi^{0}~(\%)$ &  $n\pi^{+}\pi^{-}\pi^{+}~(\%)$ & $nK^{-}\pi^{+}\pi^{+}~(\%)$  \\
			\hline
			No extra charged track & 1.1 & 1.1 & 1.1 \\
			Tracking          & 1.0 & 3.0 & 3.0 \\
			PID               & 1.0 & 3.0 & 3.0   \\
			$\pi^{0}$ reconstruction   & 1.0 & $-$ & $-$ \\
			Background veto   & 1.6 & 1.0 & $-$ \\
			$M_{\rm miss}$ fit & 1.3 & 2.2 & 0.9  \\
			ST $\bar{\Lambda}_{c}^{-}$ yield & 0.1 & 0.2 & 0.2 \\
			MC model        & 1.1 & 2.6 & $\cdots$ \\
			MC sample size     & 0.2 &  0.2  &  0.2 \\
			\hline
			Total             &3.1 & 5.6 & 4.5       \\
			\hline \hline
		\end{tabular}
		}
	\vspace{0.2cm}
		\label{tab:sys_err}
	\end{center}

\section{Summary}
In summary, by analyzing 4.5 $\ifb$ of data collected at c.m. energies between 4599.53 and 4698.82~MeV, we report for the first time the observation of $\Lambda^+_c\to n\pi^{+}\pi^{0}$, $\Lambda^+_c\to n\pi^{+}\pi^{-}\pi^{+}$ and $\Lambda^+_c\to nK^{-}\pi^{+}\pi^{+}$.
The BFs of these decays are determined to be  $\mathcal{B}(\Lambda_{c}^{+}\rightarrow n\pi^{+}\pi^{0})=(0.64\pm0.09\pm0.02)\%$, $\mathcal{B}(\Lambda_{c}^{+}\rightarrow n\pi^{+}\pi^{-}\pi^{+})=(0.45\pm0.07\pm0.03)\%$, and $\mathcal{B}(\Lambda_{c}^{+}\rightarrow nK^{-}\pi^{+}\pi^{+})=(1.90\pm0.08\pm0.09)\%$, where the first uncertainties are statistical and the second systematic.
These observations are important additions to our knowledge of  $\Lambda_c^+$ decays. Comparison of these results to those of  decays involving protons provides crucial inputs for understanding the mechanisms in the charmed baryon decays under the SU(3) flavor symmetry.
Taking $\mathcal{B}(\Lambda_c^+ \to p \pi^- \pi^+) =(0.461\pm0.028)\%$ from the PDG~\cite{pdg2021}, we can calculate $\mathcal{B}(\Lambda_c^+ \to p \pi^- \pi^+)/\mathcal{B}(\Lambda_c^+ \to n \pi^0 \pi^+) = 0.72\pm0.11$.
This result provides useful input to test of isospin symmetry in the charm baryon sector.
Taking $\mathcal{B}(\Lambda_c^+ \to n \pi^+) =(6.6\pm1.3)\times10^{-4}$~\cite{BESIII:npi}, the ratio $\mathcal{B}(\Lambda_c^+ \to n \pi^+ \pi^0)/\mathcal{B}(\Lambda_c^+ \to n \pi^+)$ is calculated to be $9.7\pm2.4$, indicating an order-of-magnitude difference in the rates of the two decays. 
This ratio is greater than $\mathcal{B}(\Lambda_c^+ \to p K_{S}^{0}\pi^0)/\mathcal{B}(\Lambda_c^+ \to p K_{S}^{0})=(1.24\pm0.10)$.
To further understand this behavior, amplitude analysis will be needed to decouple the intermediate resonances contributions for $\Lambda_c^+ \to n \pi^+ \pi^0$ and $\Lambda_c^+ \to p K_{S}^{0}\pi^0$ .
The ratio of $\mathcal{B}(\Lambda_c^+ \to n \pi^+\pi^-\pi^+)/\mathcal{B}(\Lambda_c^+ \to n K^-\pi^+\pi^+)=(0.24\pm0.04)$ which is consistent with the ratio of Cabibbo-Kobayashi-Maskawa matrix elements $|V_{cd}|/|V_{cs}|=(0.224\pm0.005)$ offers a new constraint on the CS and CF decay dynamics.
The results from this analysis provide an essential input for the phenomenological studies on the underlying dynamics of charmed bayond decays.

\acknowledgments{
The BESIII collaboration thanks the staff of BEPCII and the IHEP computing center for their strong support. 
The authors are grateful to Fusheng Yu for enlightening discussions. 
}

\begin{appendices}
	\renewcommand\thefigure{\Alph{section}\arabic{figure}}
	\renewcommand\thetable{\Alph{section}\arabic{table}}
	\section{}
	\setcounter{figure}{0}
	Figures~\ref{fig:topo_n3pi} and \ref{fig:topo_nkpipi} show the topological diagrams of $\Lambda_c^+ \to n\pi^+\pi^-\pi^+$ and $\Lambda_c^+ \to n K^-\pi^+\pi^+$, respectively.

	\begin{figure*}[htbp]
		\begin{center}
			\centering
			\subfigure[$T$]
			{
				\includegraphics[width=0.35\textwidth]{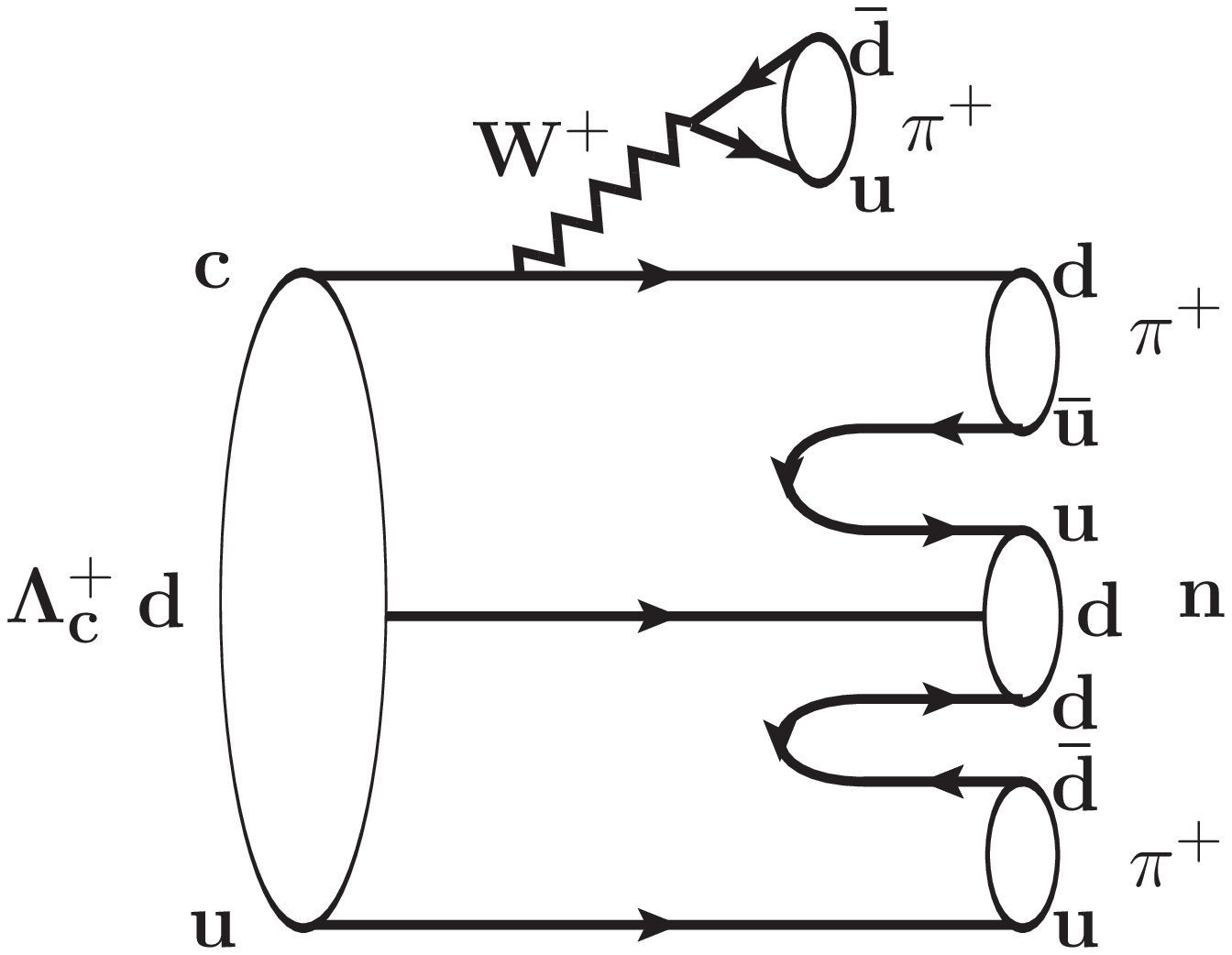}
			}
			\subfigure[$C$]
			{
				\includegraphics[width=0.35\textwidth]{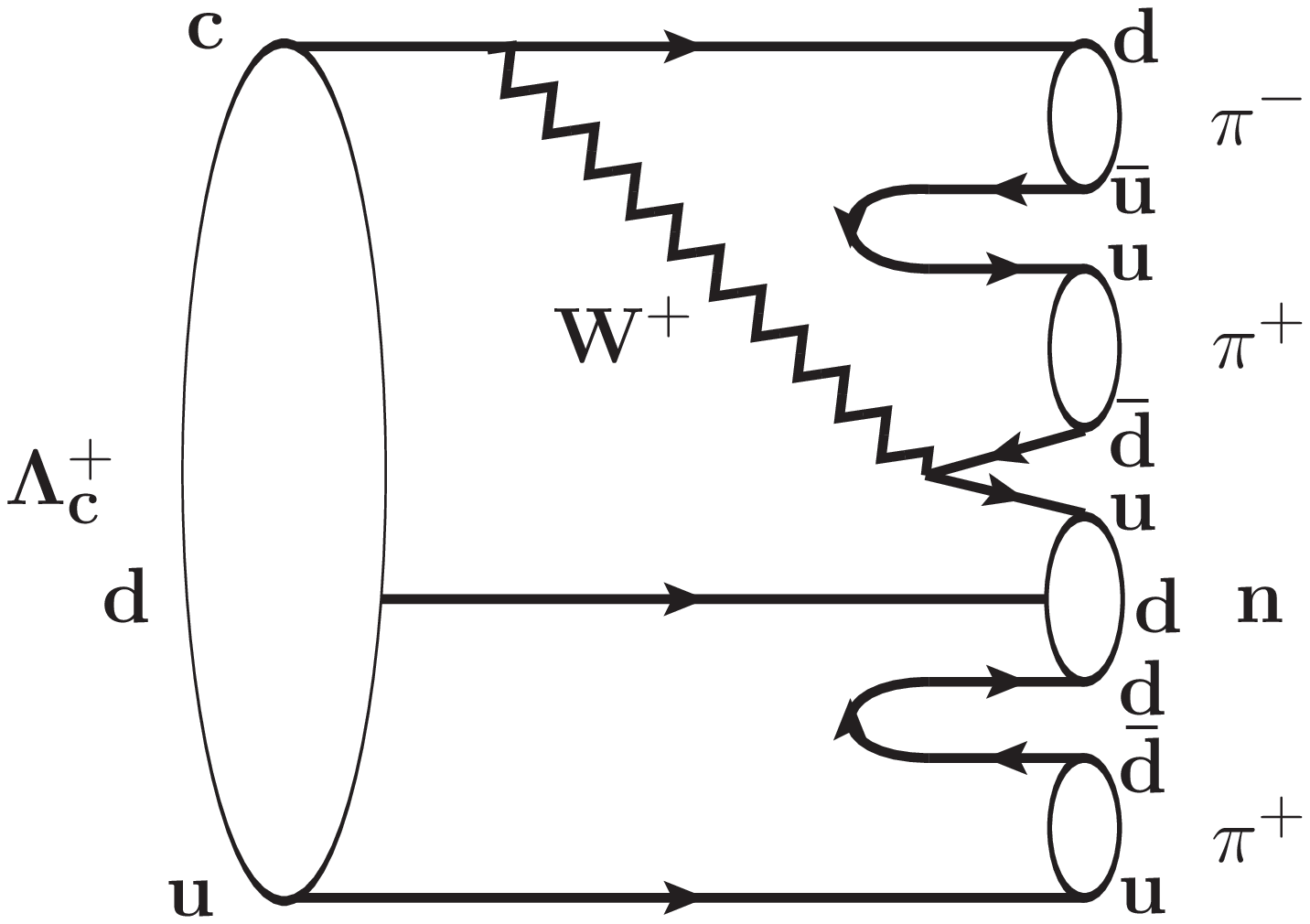}
			}
			\subfigure[$C^{\prime}$]
			{
				\includegraphics[width=0.35\textwidth]{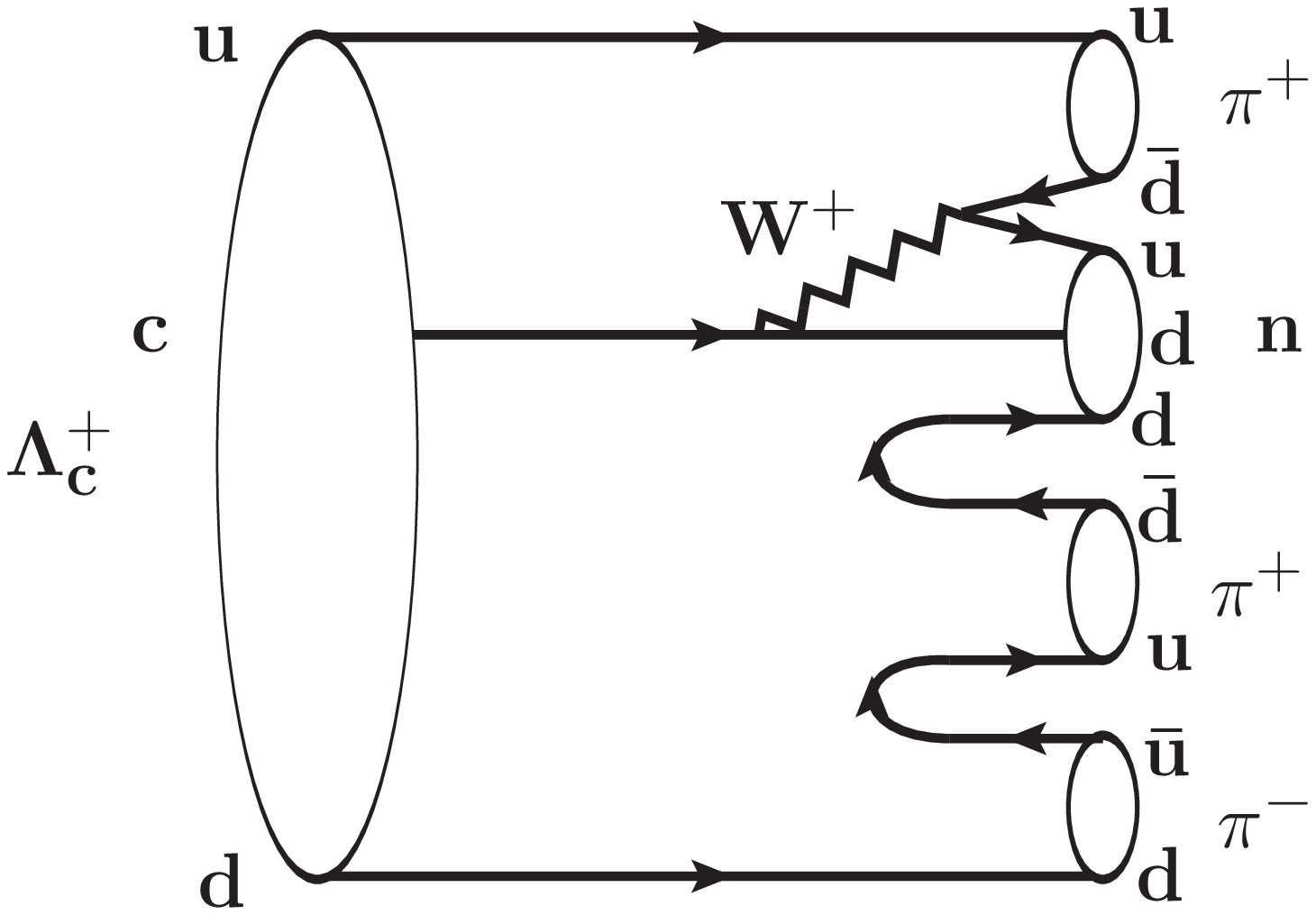}
			}
			\subfigure[$E$]
			{
				\includegraphics[width=0.35\textwidth]{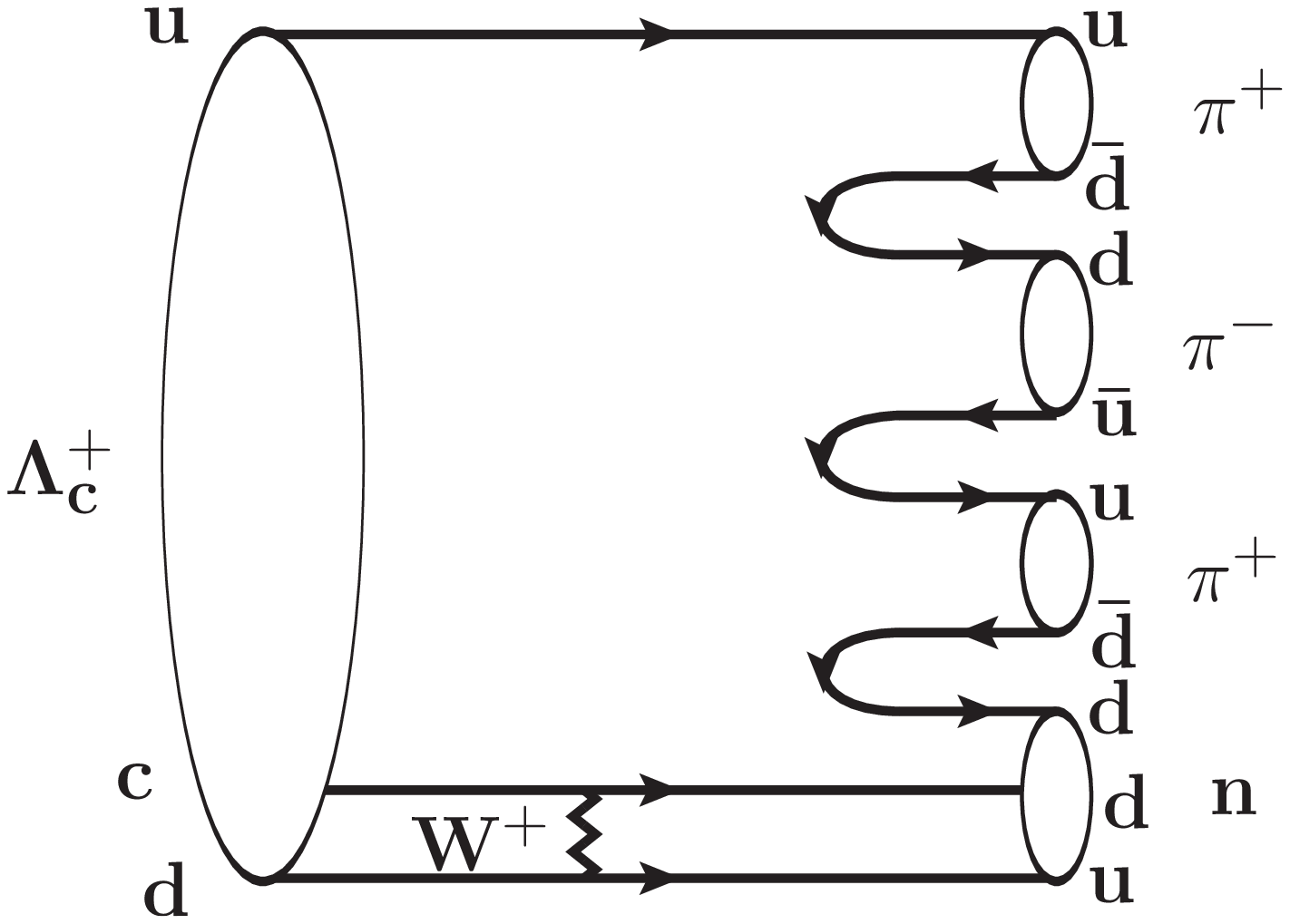}
			}
			\caption{Topological diagrams of $\Lambda_c^+ \to n\pi^+\pi^-\pi^+$ via (a) external $W$-emission $T$, (b) internal $W$-emission $C$, (c) inner $W$-emission $C^{\prime}$, and (d) $W$-exchange diagram $E$.
			}
			\label{fig:topo_n3pi}
		\end{center}
	\end{figure*}
	\begin{figure*}[htbp]
		\begin{center}
			\centering
			\subfigure[$T$]
			{
				\includegraphics[width=0.35\textwidth]{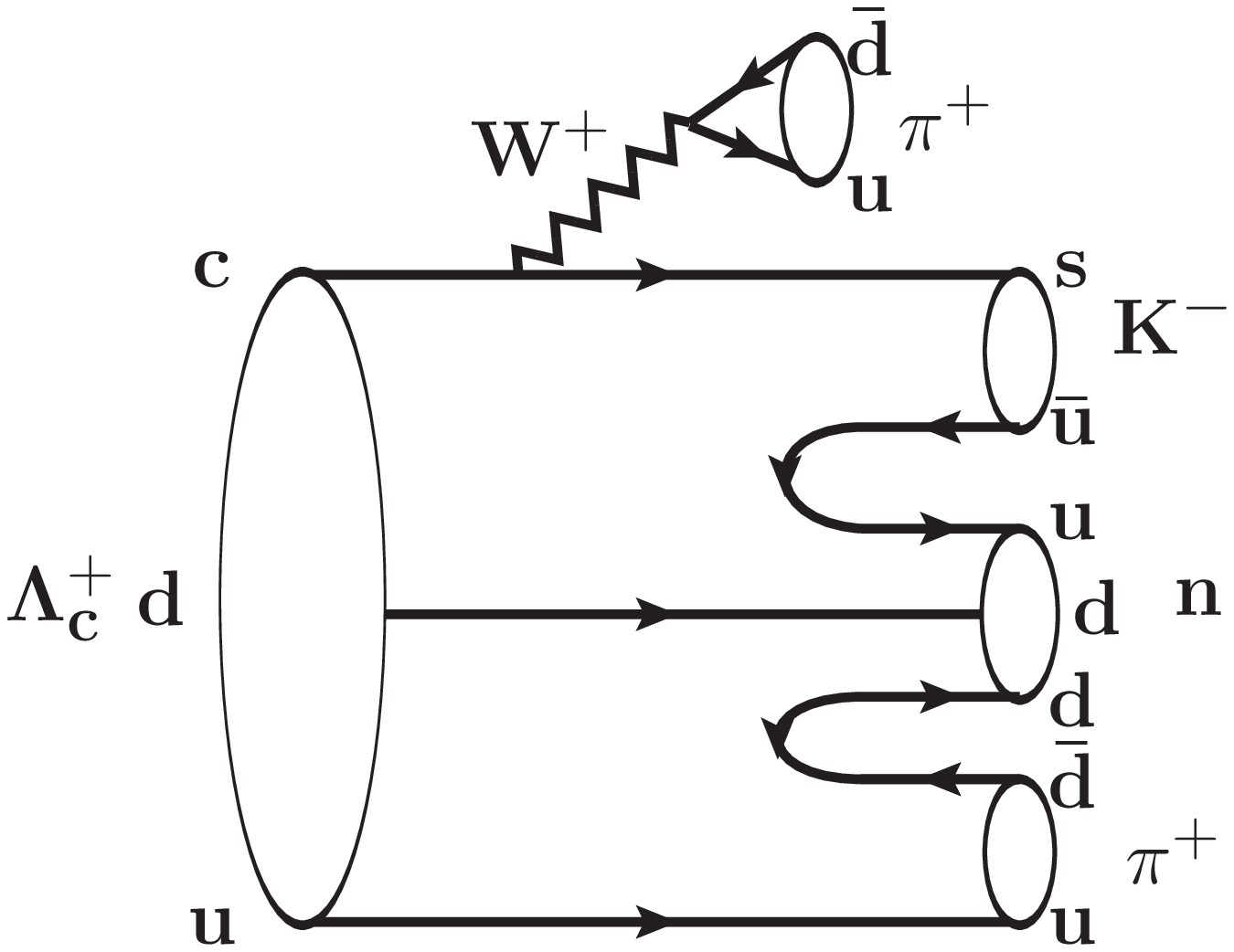}
			}
			\subfigure[$C$]
			{
				\includegraphics[width=0.35\textwidth]{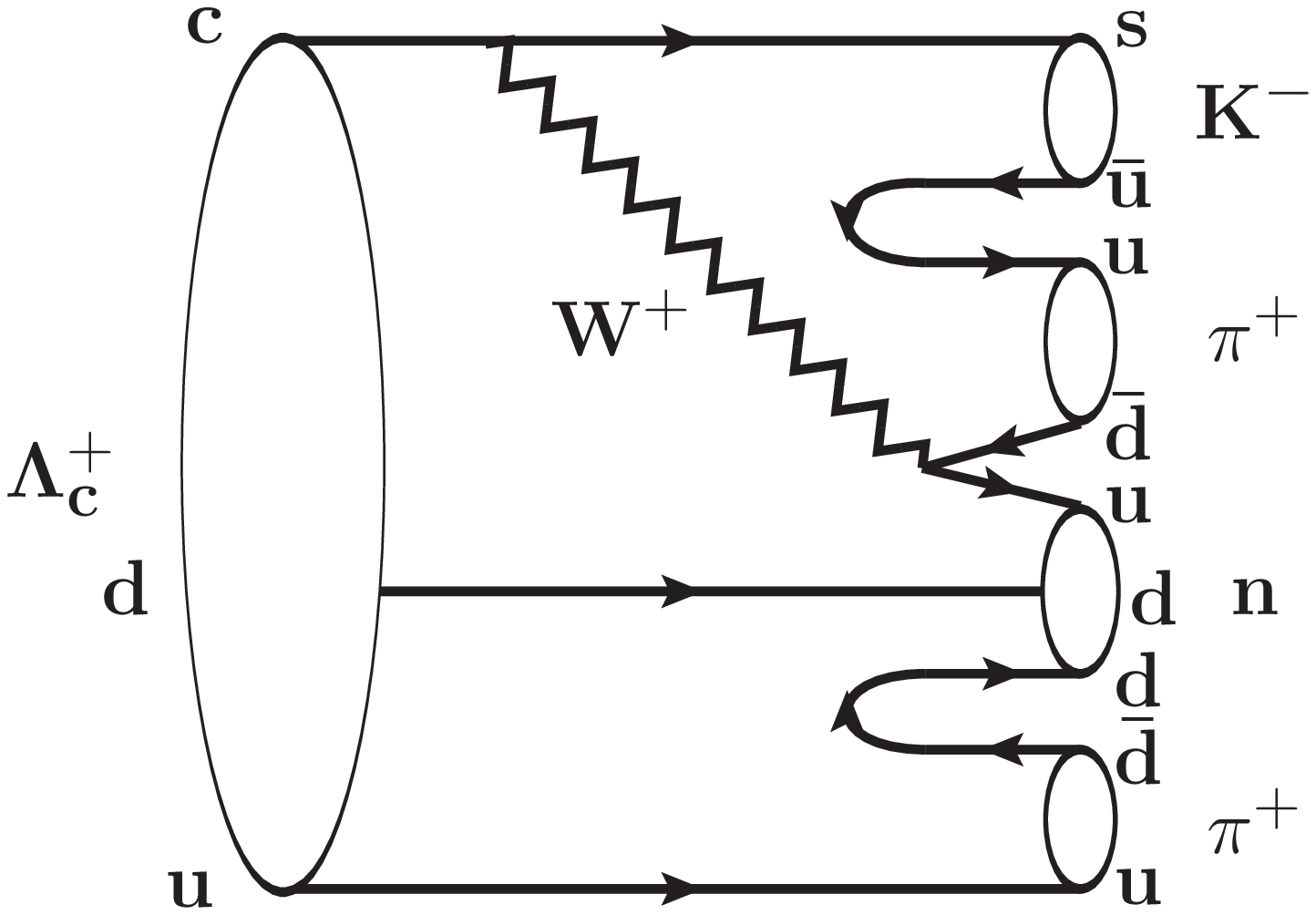}
			}
			\subfigure[$E$]
			{
				\includegraphics[width=0.35\textwidth]{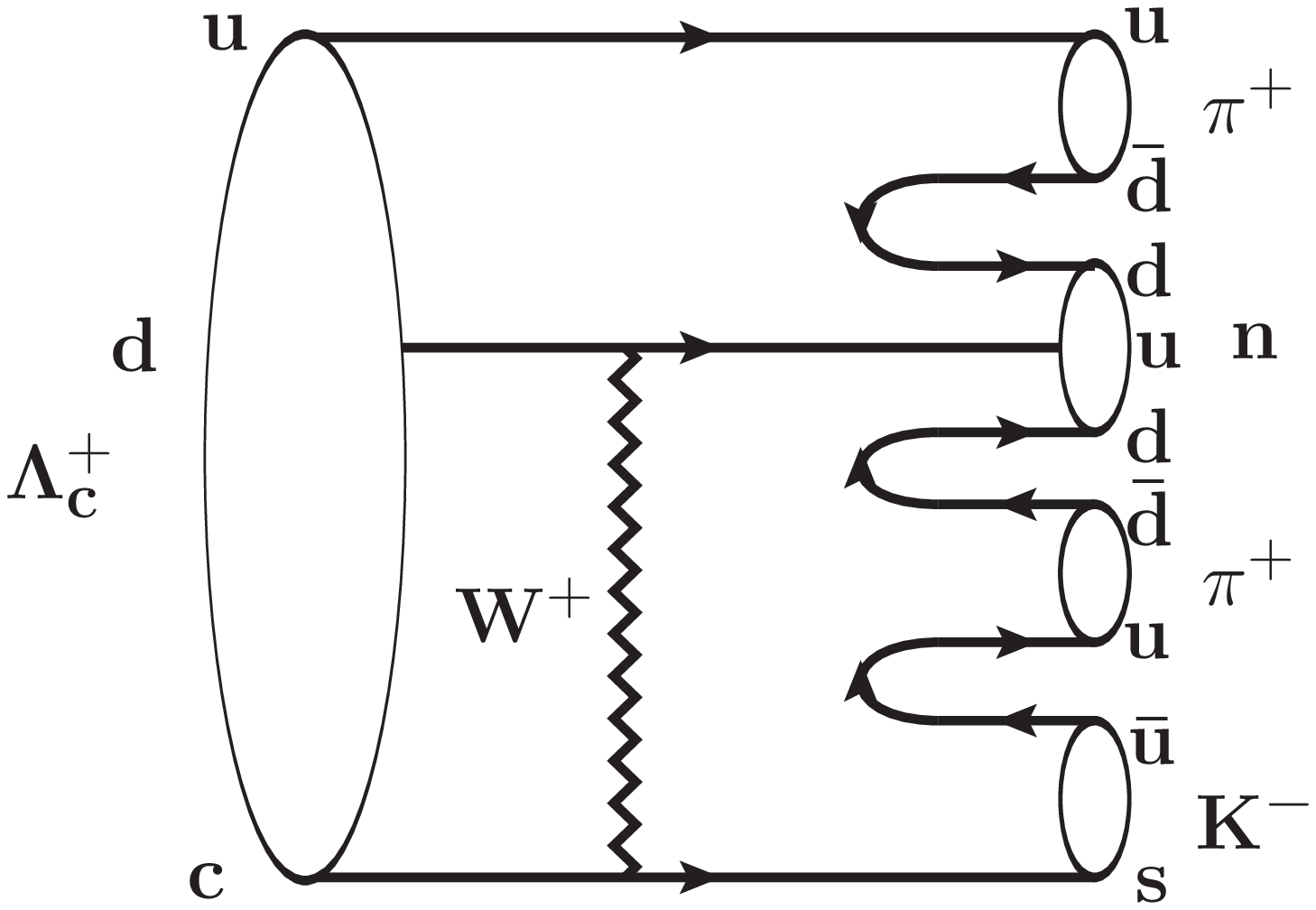}
			}
			\caption{Topological diagrams of $\Lambda_c^+ \to n K^-\pi^+\pi^+$ via (a) external $W$-emission $T$, (b) internal $W$-emission $C$, (c) $W$-exchange diagram $E$.
			}
			\label{fig:topo_nkpipi}
		\end{center}
	\end{figure*}

	\section{}
	\setcounter{figure}{0}
		Figures \ref{fig:ST_datafit_4600}$-$\ref{fig:ST_datafit_4700} show the fits for the $M_{\rm BC}$ distributions of the ST $\bar \Lambda^-_c$ candidates for various tag modes from the data samples at $\sqrt s =$ 4599.53, 4611.86, 4628.00, 4640.91, 4661.24, and 4698.82 MeV, respectively.
		
		Tables \ref{tab:4600}$-$\ref{tab:4700} show the ST yields, ST and DT efficiencies for various tag modes from the data samples at $\sqrt s =$ 4599.53, 4611.86, 4628.00, 4640.91, 4661.24, and 4698.82 MeV, respectively. 
		
			\begin{figure*}[htbp]
				\centering
				\includegraphics[width=0.6\textwidth]{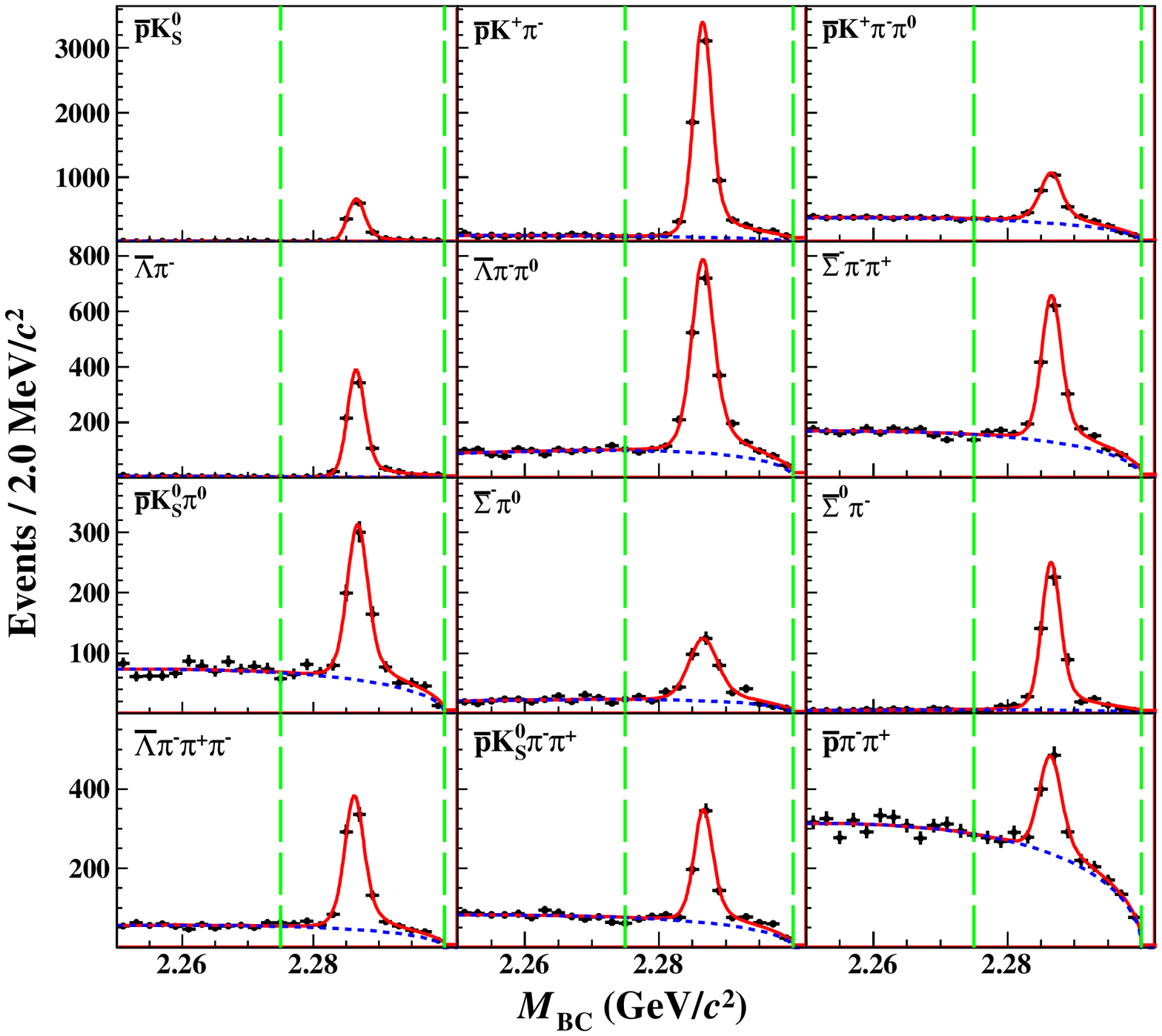}
				\caption{ The $M_{\rm BC}$ distributions of the ST $\bar \Lambda^-_c$ candidates of various tag modes for the data sample at $\sqrt{s}=4599.53~\mathrm{MeV}$.  The points with error bars represent data. The (red) solid curves indicate the fit results and the (blue) dashed curves describe the background shapes. The ranges between (green) dashed lines are the signal regions.} 
				\label{fig:ST_datafit_4600}
			\end{figure*}

			\begin{figure*}[htbp]
				\centering
				\includegraphics[width=0.6\textwidth]{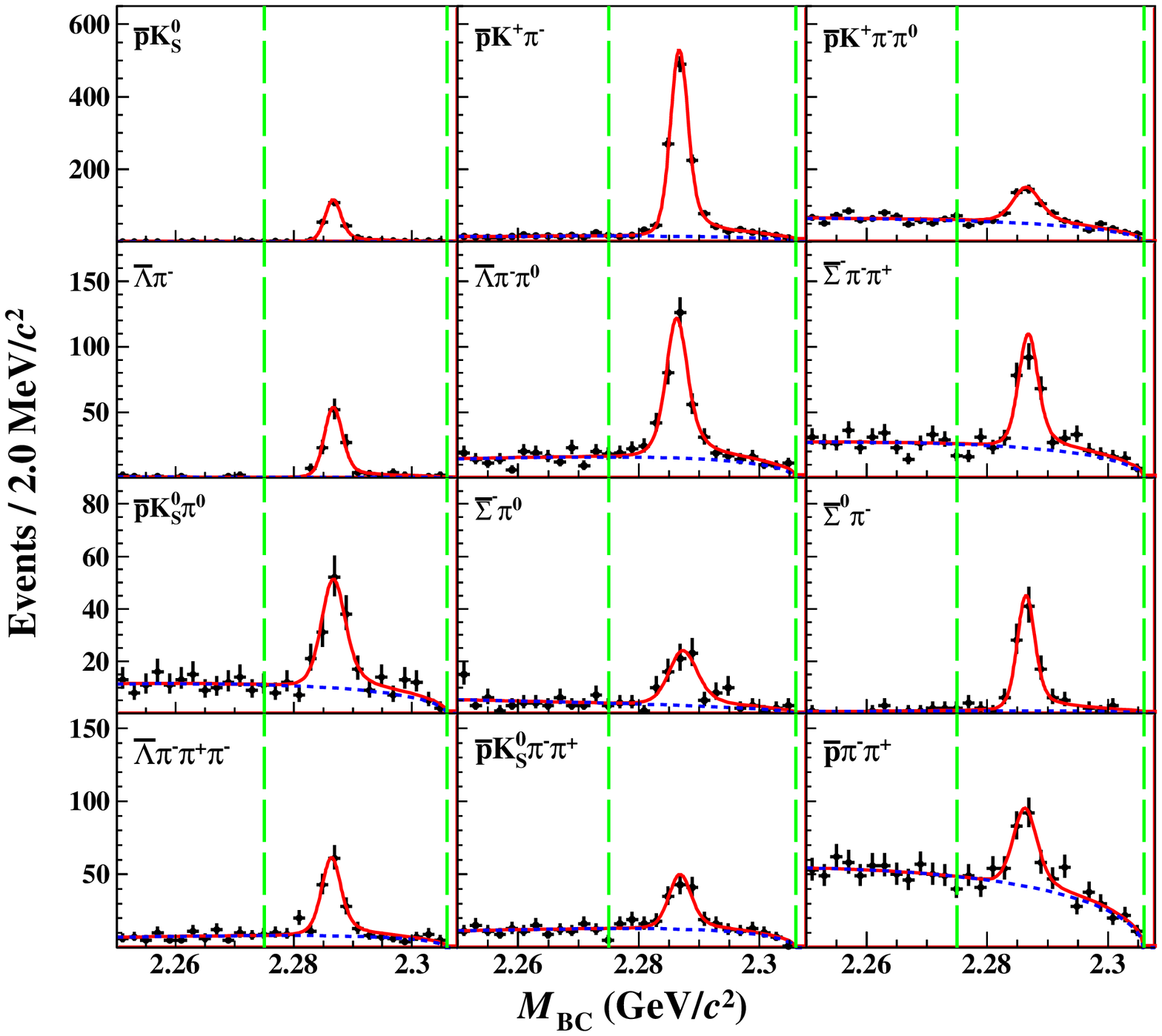}
				\caption{ The $M_{\rm BC}$ distributions of the ST $\bar \Lambda^-_c$ candidates of various tag modes for the data sample at $\sqrt{s}=4611.86~\mathrm{MeV}$.  The points with error bars represent data. The (red) solid curves indicate the fit results and the (blue) dashed curves describe the background shapes. The ranges between (green) dashed lines are the signal regions.} 
				\label{fig:ST_datafit_4612}
			\end{figure*}
			
			\begin{figure*}[htbp]
				\centering
				\includegraphics[width=0.6\textwidth]{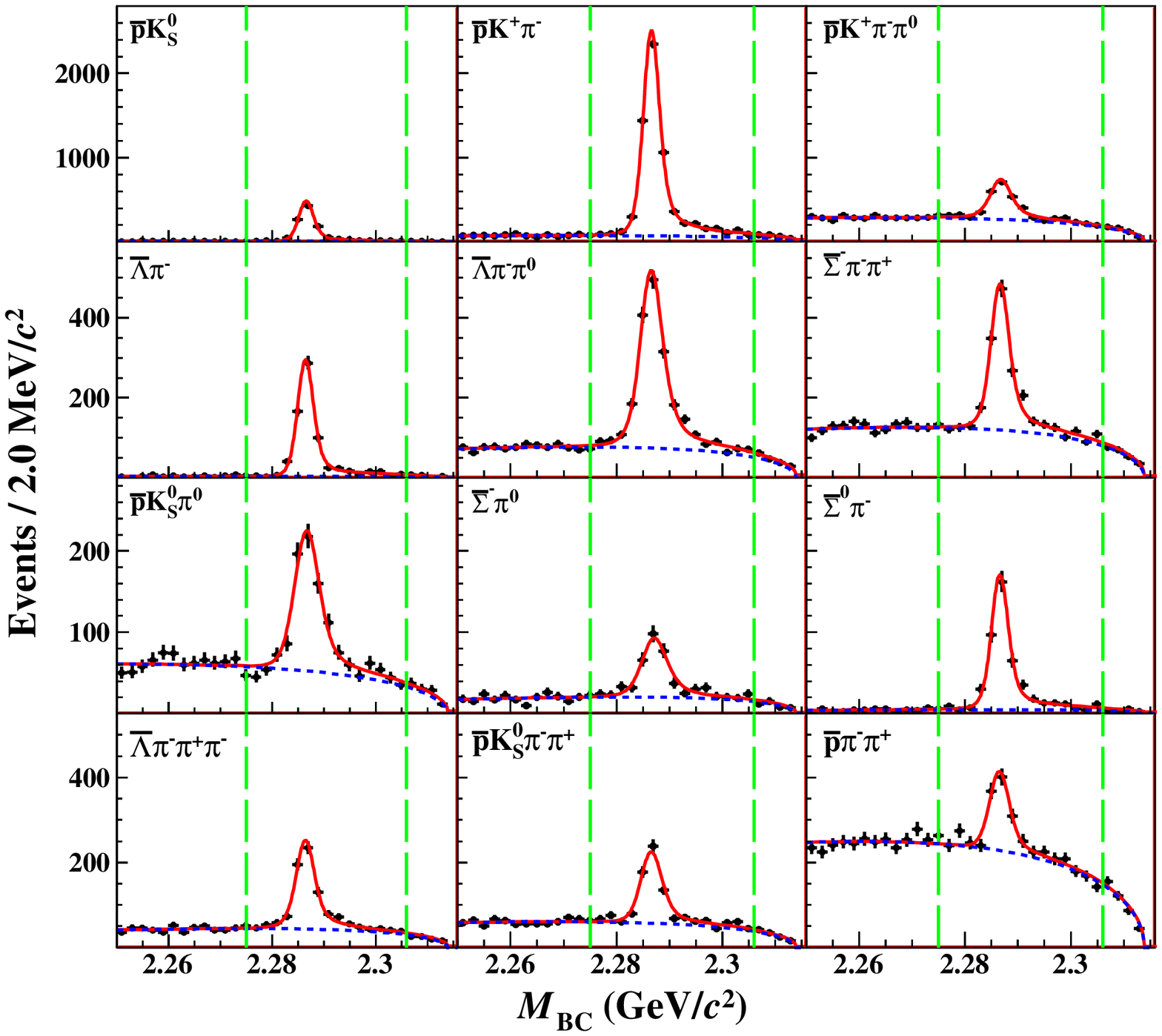}
				\caption{ The $M_{\rm BC}$ distributions of the ST $\bar \Lambda^-_c$ candidates of various tag modes for the data sample at $\sqrt{s}=4628.00~\mathrm{MeV}$.  The points with error bars represent data. The (red) solid curves indicate the fit results and the (blue) dashed curves describe the background shapes. The ranges between (green) dashed lines are the signal regions.} 
				\label{fig:ST_datafit_4626}
			\end{figure*}
			
			\begin{figure*}[htbp]
				\centering
				\includegraphics[width=0.6\textwidth]{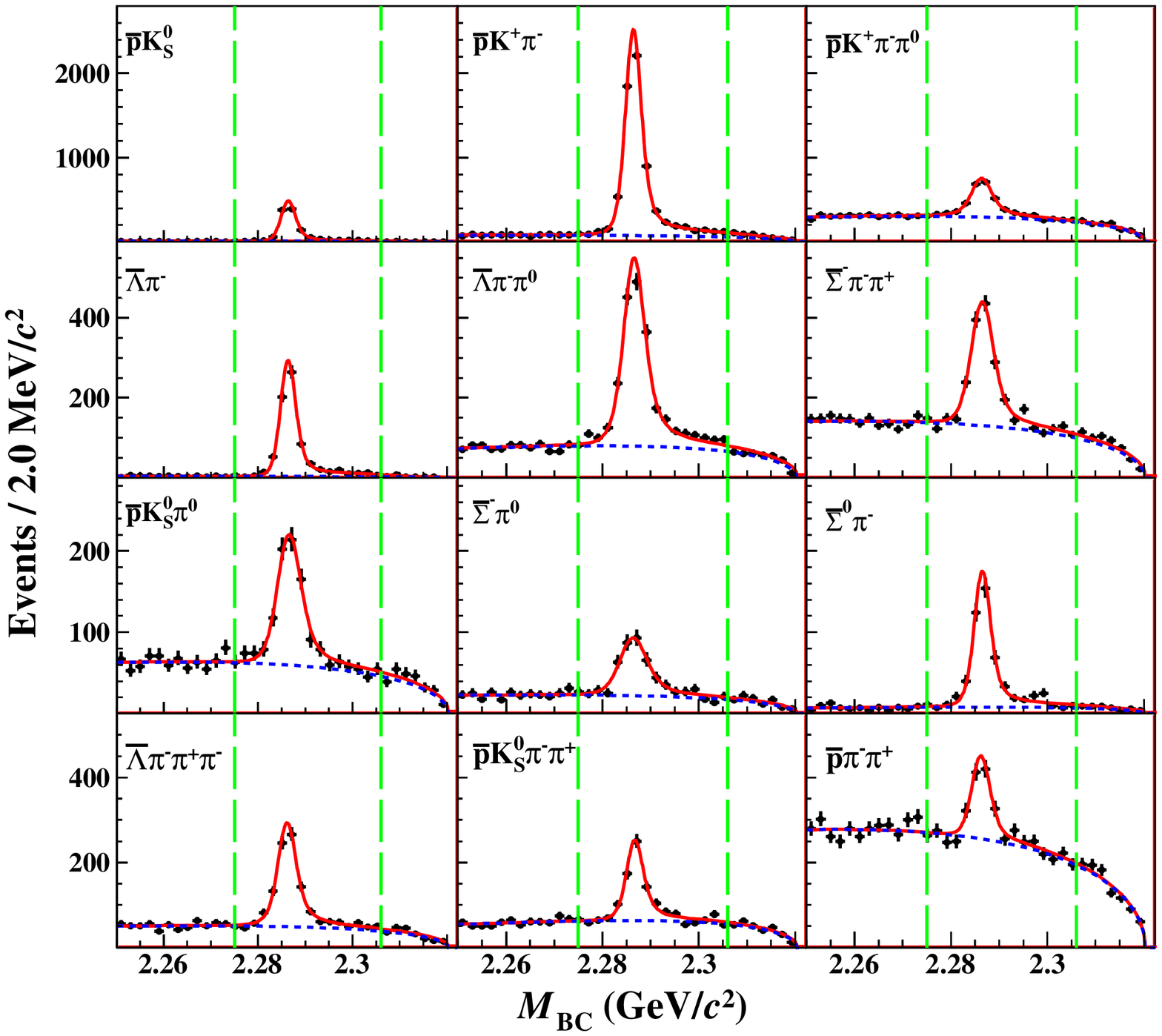}
				\caption{ The $M_{\rm BC}$ distributions of the ST $\bar \Lambda^-_c$ candidates of various tag modes for the data sample at $\sqrt{s}=4640.91~\mathrm{MeV}$.  The points with error bars represent data. The (red) solid curves indicate the fit results and the (blue) dashed curves describe the background shapes. The ranges between (green) dashed lines are the signal regions.} 
				\label{fig:ST_datafit_4640}
			\end{figure*}

			\begin{figure*}[htbp]
				\centering
				\includegraphics[width=0.6\textwidth]{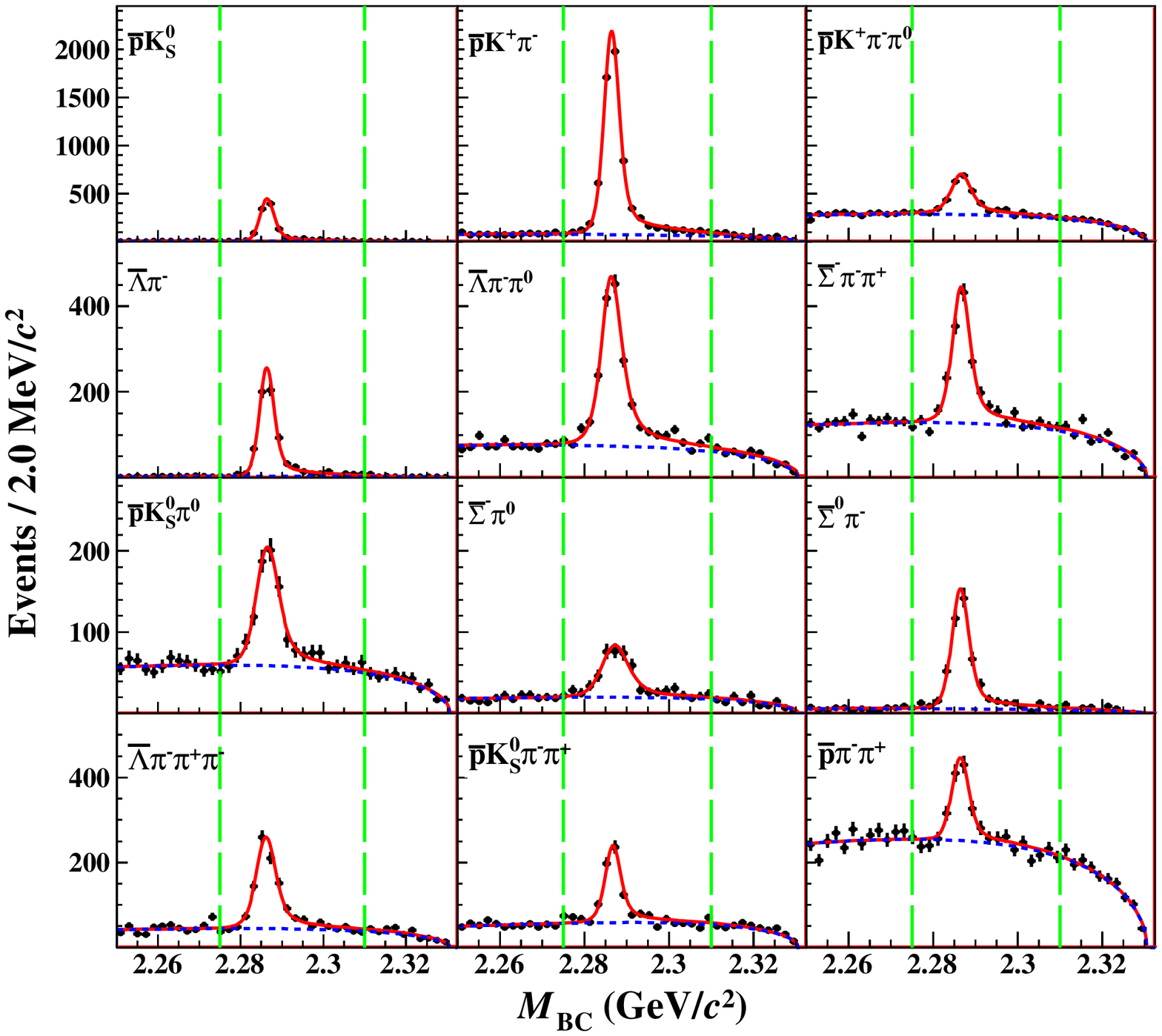}
				\caption{ The $M_{\rm BC}$ distributions of the ST $\bar \Lambda^-_c$ candidates of various tag modes for the data sample at $\sqrt{s}=4661.24~\mathrm{MeV}$.  The points with error bars represent data. The (red) solid curves indicate the fit results and the (blue) dashed curves describe the background shapes. The ranges between (green) dashed lines are the signal regions.} 
				\label{fig:ST_datafit_4660}
			\end{figure*}
			
			\begin{figure*}[htbp]
				\centering
				\includegraphics[width=0.6\textwidth]{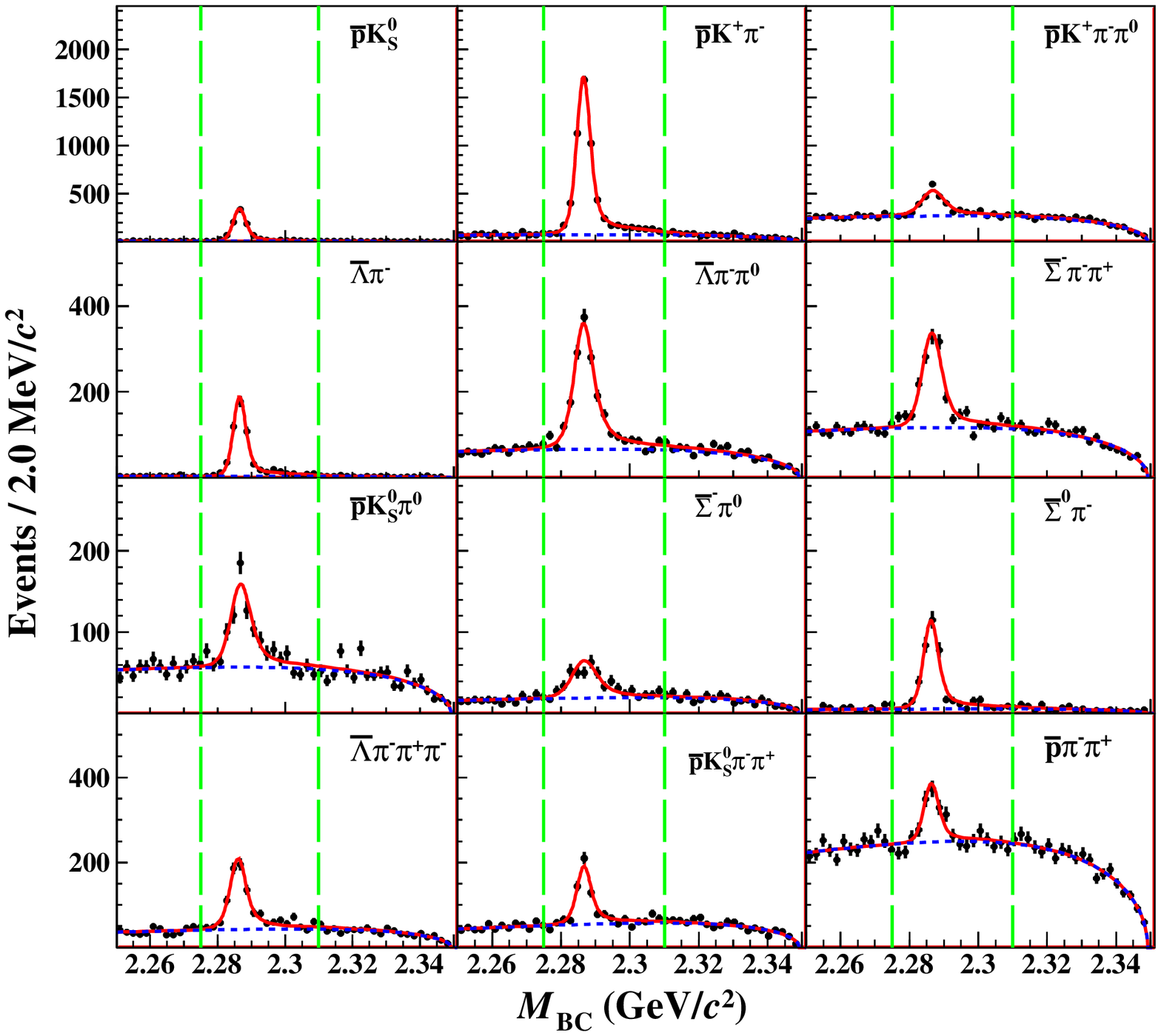}
				\caption{ The $M_{\rm BC}$ distributions of the ST $\bar \Lambda^-_c$ candidates of various tag modes for the data sample at $\sqrt{s}=4698.82~\mathrm{MeV}$.  The points with error bars represent data. The (red) solid curves indicate the fit results and the (blue) dashed curves describe the background shapes. The ranges between (green) dashed lines are the signal regions.} 
				\label{fig:ST_datafit_4700}
			\end{figure*}
			
			\setcounter{table}{0}
			\begin{table*}[htbp]
				\begin{center}
					\caption{ST yields, ST and DT efficiencies of various tag modes for the data sample at $\sqrt{s}=4599.53\mev$. The uncertainties are statistical only. The quoted efficiencies do not include any subdecay BFs. Entries of ``$\cdots$" are for the cases where knowledge of the DT efficiencies are not required in the analysis.}
					\begin{tabular}{l|c|c|c|c|c}
						\hline \hline
						Tag mode  & $N_{i}^{\rm ST}$ & $\varepsilon_{i}^{\rm ST}(\%) $ & $\varepsilon_{i}^{\rm DT}(n\pi^+\pi^0)(\%)$ & $\varepsilon_{i}^{\rm DT}(n\pi^{+}\pi^{-}\pi^{+})(\%)$   & $\varepsilon_{i}^{\rm DT}(nK^{-}\pi^{+}\pi^{-})(\%)$ \\ \hline
						$\textbf{$\Modea$}$   & $1277\pm36$ & $56.1$ & $14.27$ & $13.49$ & $20.87$ \\
						$\textbf{$\Modeb$}$   & $6806\pm91$ & $51.5$ & $11.25$ & $11.53$ & $17.64$ \\
						$\textbf{$\Modec$}$   & $606\pm34$  & $23.0$ & $5.17$  & $5.25$  & $7.34$  \\
						$\textbf{$\Moded$}$   & $613\pm34$  & $23.5$ & $4.51$  & $4.76$  & $6.89$  \\
						$\textbf{$\Modee$}$   & $2197\pm78$ & $20.6$ &    $\cdots$    & $4.57$  & $6.83$  \\
						$\textbf{$\Modeaa$}$  & $757\pm28$  & $48.4$ & $11.62$ & $10.9$   & $17.08$ \\
						$\textbf{$\Modebb$}$  & $1742\pm56$ & $21.6$ & $4.52$  & $4.43$  & $6.42$  \\
						$\textbf{$\Modedd$}$  & $769\pm36$  & $15.6$ & $2.86$  & $2.90$  & $4.17$  \\
						$\textbf{$\Modeaaa$}$ & $520\pm26$  & $29.4$ & $7.30$  & $6.96$  & $11.20$  \\
						$\textbf{$\Modeccc$}$ & $320\pm25$  & $23.7$ & $6.47$  & $6.05$  & $8.38$  \\
						$\textbf{$\Modeddd$}$ & $1186\pm49$ & $25.4$ & $5.84$  & $5.82$  & $8.81$  \\
						$\textbf{$\Modef$}$   & $598\pm47$  & $64.3$ &    $\cdots$    &    $\cdots$    & $22.04$ \\
						\hline  \hline
					\end{tabular}
					\label{tab:4600}
				\end{center}
			\end{table*}
			
			\begin{table*}[htbp]
				\begin{center}
					\caption{ST yields, ST and DT efficiencies of various tag modes for the data sample at $\sqrt{s}=4611.86\mev$. The uncertainties are statistical only. The quoted efficiencies do not include any subdecay BFs. Entries of ``$\cdots$" are for the cases where knowledge of the DT efficiencies are not required in the analysis.}
					\begin{tabular}{l|c|c|c|c|c}
						\hline \hline
						Tag mode  & $N_{i}^{\rm ST}$ & $\varepsilon_{i}^{\rm ST}(\%) $ & $\varepsilon_{i}^{\rm DT}(n\pi^+\pi^0)(\%)$ & $\varepsilon_{i}^{\rm DT}(n\pi^{+}\pi^{-}\pi^{+})(\%)$   & $\varepsilon_{i}^{\rm DT}(nK^{-}\pi^{+}\pi^{-})(\%)$ \\ \hline
						$\textbf{$\Modea$}$   & $239\pm16$  & $53.7$ & $13.89$ & $12.71$ & $20.09$ \\
						$\textbf{$\Modeb$}$   & $1166\pm39$ & $51.0$ & $11.34$ & $11.27$ & $16.98$ \\
						$\textbf{$\Modec$}$   & $127\pm17$  & $22.2$ & $5.04$  & $4.91$  & $7.20$  \\
						$\textbf{$\Moded$}$   & $106\pm16$  & $21.9$ & $4.35$  & $4.33$  & $6.04$  \\
						$\textbf{$\Modee$}$   & $364\pm34$  & $19.9$ &    $\cdots$    & $4.40$  & $6.25$  \\
						$\textbf{$\Modeaa$}$  & $123\pm11$  & $46.9$ & $11.29$ & $10.22$  & $15.27$ \\
						$\textbf{$\Modebb$}$  & $302\pm23$  & $19.8$ & $4.34$  & $4.17$  & $6.08$  \\
						$\textbf{$\Modedd$}$  & $139\pm15$  & $13.6$ & $2.68$  & $2.75$  & $3.69$  \\
						$\textbf{$\Modeaaa$}$ & $102\pm13$  & $26.6$ & $6.89$  & $6.53$  & $8.65$  \\
						$\textbf{$\Modeccc$}$ & $73\pm10$   & $22.6$ & $6.47$  & $5.9$   & $8.46$  \\
						$\textbf{$\Modeddd$}$ & $218\pm22$  & $25.5$ & $5.74$  & $5.63$  & $8.47$  \\
						$\textbf{$\Modef$}$   & $155\pm22$  & $71.4$ &    $\cdots$    &    $\cdots$    & $21.82$ \\
						\hline  \hline
					\end{tabular}
					\label{tab:4612}
				\end{center}
			\end{table*}

			\begin{table*}[htbp]
				\begin{center}
					\caption{ST yields, ST and DT efficiencies of various tag modes for the data sample at $\sqrt{s}=4628.00\mev$. The uncertainties are statistical only. The quoted efficiencies do not include any subdecay BFs.Entries of ``$\cdots$" are for the cases where knowledge of the DT efficiencies are not required in the analysis.}
					\begin{tabular}{l|c|c|c|c|c}
						\hline \hline
						Tag mode  & $N_{i}^{\rm ST}$ & $\varepsilon_{i}^{\rm ST}(\%) $ & $\varepsilon_{i}^{\rm DT}(n\pi^+\pi^0)(\%)$ & $\varepsilon_{i}^{\rm DT}(n\pi^{+}\pi^{-}\pi^{+})(\%)$   & $\varepsilon_{i}^{\rm DT}(nK^{-}\pi^{+}\pi^{-})(\%)$ \\ \hline
						$\textbf{$\Modea$}$   & $1054\pm35$ & $51.8$ & $13.27$ & $12.17$ & $18.98$ \\
						$\textbf{$\Modeb$}$   & $5886\pm39$ & $49.2$ & $10.98$ & $10.97$ & $16.77$ \\
						$\textbf{$\Modec$}$   & $616\pm36$  & $20.7$ & $4.94$  & $4.79$  & $6.73$  \\
						$\textbf{$\Moded$}$   & $510\pm32$  & $20.6$ & $4.21$  & $4.17$  & $6.01$  \\
						$\textbf{$\Modee$}$   & $1589\pm69$ & $18.7$ &    $\cdots$    & $4.32$  & $6.95$  \\
						$\textbf{$\Modeaa$}$  & $675\pm28$  & $43.2$ & $10.69$ & $9.84$   & $14.72$ \\
						$\textbf{$\Modebb$}$  & $1454\pm54$ & $19.1$ & $4.18$  & $3.98$  & $5.90$  \\
						$\textbf{$\Modedd$}$  & $587\pm33$  & $13.6$ & $2.70$  & $2.65$  & $3.67$  \\
						$\textbf{$\Modeaaa$}$ & $413\pm23$  & $27.2$ & $6.62$  & $6.24$  & $8.41$  \\
						$\textbf{$\Modeccc$}$ & $263\pm23$  & $23.4$ & $6.20$  & $5.65$  & $9.10$  \\
						$\textbf{$\Modeddd$}$ & $994\pm20$  & $23.6$ & $5.54$  & $5.47$  & $8.24$  \\
						$\textbf{$\Modef$}$   & $517\pm45$  & $61.6$ &    $\cdots$    &    $\cdots$    & $19.75$ \\
						\hline  \hline
					\end{tabular}
					\label{tab:4626}
				\end{center}
			\end{table*}
			
			\begin{table*}[htbp]
				\begin{center}
					\caption{ST yields, ST and DT efficiencies of various tag modes for the data sample at $\sqrt{s}=4640.91\mev$. The uncertainties are statistical only. The quoted efficiencies do not include any subdecay BFs. Entries of ``$\cdots$" are for the cases where knowledge of the DT efficiencies are not required in the analysis.}
					\begin{tabular}{l|c|c|c|c|c}
						\hline \hline
						Tag mode  & $N_{i}^{\rm ST}$ & $\varepsilon_{i}^{\rm ST}(\%) $ & $\varepsilon_{i}^{\rm DT}(n\pi^+\pi^0)(\%)$ & $\varepsilon_{i}^{\rm DT}(n\pi^{+}\pi^{-}\pi^{+})(\%)$   & $\varepsilon_{i}^{\rm DT}(nK^{-}\pi^{+}\pi^{-})(\%)$ \\ \hline
						$\textbf{$\Modea$}$   & $1107\pm36$ & $50.7$ & $13.14$ & $12.08$ & $19.12$ \\
						$\textbf{$\Modeb$}$   & $6250\pm89$ & $48.5$ & $10.91$ & $10.86$ & $16.83$ \\
						$\textbf{$\Modec$}$   & $599\pm36$  & $20.7$ & $4.82$  & $4.77$  & $7.02$  \\
						$\textbf{$\Moded$}$   & $522\pm33$  & $20.8$ & $4.21$  & $4.15$  & $6.50$  \\
						$\textbf{$\Modee$}$   & $1632\pm70$ & $18.1$ &    $\cdots$    & $4.23$  & $6.62$  \\
						$\textbf{$\Modeaa$}$  & $705\pm29$  & $42.7$ & $10.51$ & $9.55$  & $14.24$ \\
						$\textbf{$\Modebb$}$  & $1613\pm54$ & $19.1$ & $4.14$  & $3.95$  & $5.46$  \\
						$\textbf{$\Modedd$}$  & $745\pm36$  & $14.2$ & $2.70$  & $2.67$  & $4.20$  \\
						$\textbf{$\Modeaaa$}$ & $445\pm25$  & $26.2$ & $6.43$  & $6.09$  & $8.97$  \\
						$\textbf{$\Modeccc$}$ & $298\pm24$  & $24.6$ & $6.01$  & $5.52$  & $7.70$  \\
						$\textbf{$\Modeddd$}$ & $1077\pm49$ & $23.4$ & $5.43$  & $5.38$  & $8.01$  \\
						$\textbf{$\Modef$}$   & $552\pm47$  & $59.7$ &    $\cdots$    &    $\cdots$    & $22.97$ \\
						\hline  \hline
					\end{tabular}
					\label{tab:4640}
				\end{center}
			\end{table*}
			
			\begin{table*}[htbp]
				\begin{center}
					\caption{ST yields, ST and DT efficiencies of various tag modes for the data sample at $\sqrt{s}=4661.24\mev$. The uncertainties are statistical only. The quoted efficiencies do not include any subdecay BFs. Entries of ``$\cdots$" are for the cases where knowledge of the DT efficiencies are not required in the analysis.}
					\begin{tabular}{l|c|c|c|c|c}
						\hline \hline
						Tag mode  & $N_{i}^{\rm ST}$ & $\varepsilon_{i}^{\rm ST}(\%) $ & $\varepsilon_{i}^{\rm DT}(n\pi^+\pi^0)(\%)$ & $\varepsilon_{i}^{\rm DT}(n\pi^{+}\pi^{-}\pi^{+})(\%)$   & $\varepsilon_{i}^{\rm DT}(nK^{-}\pi^{+}\pi^{-})(\%)$ \\ \hline
						$\textbf{$\Modea$}$   & $1119\pm35$ & $49.6$ & $12.82$ & $11.78$ & $18.58$ \\
						$\textbf{$\Modeb$}$   & $5938\pm86$ & $48.2$ & $10.93$ & $10.85$ & $16.47$ \\
						$\textbf{$\Modec$}$   & $594\pm36$  & $20.1$ & $4.86$  & $4.75$  & $7.12$  \\
						$\textbf{$\Moded$}$   & $537\pm33$  & $20.2$ & $4.29$  & $4.17$  & $6.27$  \\
						$\textbf{$\Modee$}$   & $1700\pm73$ & $18.1$ &    $\cdots$    & $4.15$  & $6.36$  \\
						$\textbf{$\Modeaa$}$  & $668\pm27$  & $41.7$ & $10.15$ & $9.41$  & $14.98$ \\
						$\textbf{$\Modebb$}$  & $1491\pm51$ & $18.9$ & $4.15$  & $3.91$  & $5.91$  \\
						$\textbf{$\Modedd$}$  & $780\pm36$  & $14.1$ & $2.71$  & $2.63$  & $4.04$  \\
						$\textbf{$\Modeaaa$}$ & $454\pm25$  & $26.3$ & $6.56$  & $5.94$  & $9.20$  \\
						$\textbf{$\Modeccc$}$ & $298\pm25$  & $23.2$ & $5.96$  & $5.35$  & $8.22$  \\
						$\textbf{$\Modeddd$}$ & $1066\pm49$ & $23.2$ & $5.48$  & $5.34$  & $8.04$  \\
						$\textbf{$\Modef$}$   & $590\pm48$  & $60.2$ &    $\cdots$    &    $\cdots$    & $20.06$ \\
						\hline  \hline
					\end{tabular}
					\label{tab:4660}
				\end{center}
			\end{table*}
			
			\begin{table*}[htbp]
				\begin{center}
					\caption{ST yields, ST and DT efficiencies of various tag modes for the data sample at $\sqrt{s}=4698.82\mev$. The uncertainties are statistical only. The quoted efficiencies do not include any subdecay BFs. Entries of ``$\cdots$" are for the cases where knowledge of the DT efficiencies are not required in the analysis.}
					\begin{tabular}{l|c|c|c|c|c}
						\hline \hline
						Tag mode  & $N_{i}^{\rm ST}$ & $\varepsilon_{i}^{\rm ST}(\%) $ & $\varepsilon_{i}^{\rm DT}(n\pi^+\pi^0)(\%)$ & $\varepsilon_{i}^{\rm DT}(n\pi^{+}\pi^{-}\pi^{+})(\%)$   & $\varepsilon_{i}^{\rm DT}(nK^{-}\pi^{+}\pi^{-})(\%)$ \\ \hline
						$\textbf{$\Modea$}$   &  $958\pm33$ & $47.5$ & $12.29$ & $11.25$ & $17.86$ \\
						$\textbf{$\Modeb$}$   & $5167\pm80$ & $46.3$ & $10.62$ & $10.49$ & $16.16$ \\
						$\textbf{$\Modec$}$   &  $471\pm34$ & $18.9$ & $4.66$  & $4.52$  & $6.62$  \\
						$\textbf{$\Moded$}$   &  $462\pm31$ & $19.5$ & $4.21$  & $4.11$  & $6.33$  \\
						$\textbf{$\Modee$}$   & $1389\pm74$ & $17.5$ &   $\cdots$     & $4.02$  & $6.63$  \\
						$\textbf{$\Modeaa$}$  &  $538\pm25$ & $39.4$ & $9.80$  & $8.81$  & $13.64$ \\
						$\textbf{$\Modebb$}$  &  $1301\pm49$ &$17.9$ & $3.93$  & $3.73$  & $5.82$  \\
						$\textbf{$\Modedd$}$  &  $639\pm34$ & $14.6$ & $2.73$  & $2.63$  & $3.82$  \\
						$\textbf{$\Modeaaa$}$ &  $371\pm22$ & $24.3$ & $6.12$  & $5.75$  & $8.4$   \\
						$\textbf{$\Modeccc$}$ &  $251\pm24$ & $22.2$ & $5.62$  & $5.18$  & $7.90$  \\
						$\textbf{$\Modeddd$}$ &  $956\pm48$ & $22.2$ & $5.25$  & $5.07$  & $8.16$  \\
						$\textbf{$\Modef$}$   &  $459\pm47$ & $55.9$ &   $\cdots$     &   $\cdots$     & $19.66$ \\
						\hline  \hline
					\end{tabular}
					\label{tab:4700}
				\end{center}
			\end{table*}
	
\end{appendices}
\end{multicols}
\newpage
\vspace{-1mm}
\centerline{\rule{80mm}{0.1pt}}
\vspace{2mm}
\begin{multicols}{2}

\end{multicols}

\end{CJK*}

\end{document}